# Augmented Hybrid TV Services. Habits, Preferences, Opinions and Expectations of Spanish Consumers

Fernando Boronat, *Senior, IEEE*, Bernat Roig, Enric Sigalat, Santiago J. Flores

***Abstract*—The convergence of broadcast and broadband (i.e., hybrid) delivery, companion devices, and immersive technologies enables new forms of interactive, personalized, and accessible television (TV) services. This paper introduces the concept of Augmented Hybrid TV services and presents an exploratory user requirements and preliminary acceptance assessment for five representative use cases. The use cases were selected through an exploratory stakeholder consultation and three co-design workshops involving audiovisual professionals and participants associated with organizations supporting people with visual or hearing impairments. The use cases combine conventional TV with complementary content delivered to connected devices, including augmented and virtual reality (AR/VR) head-mounted displays. They were subsequently evaluated through a cross-sectional Computer-Assisted Web Interviewing survey of 626 Internet and connected TV users in Spain, recruited from a professionally managed, quota-based online panel. Each participant viewed standardized video demonstrations of two randomly assigned use cases and assessed their perceived usefulness, anticipated benefits and limitations, and interest in them. The results show substantial stated interest across all five use cases, ranging from 64.7% to 76.9%. Adaptive and personalized augmented subtitles received the highest level of interest and were particularly valued for their accessibility, language-learning, and personalization benefits. Immersive services involving complementary AR content, virtual screens, 3D objects, and 360° video were also evaluated positively, although participants identified device cost, ergonomics, connectivity, usability, and synchronization as anticipated limitations. The findings provide user-centered evidence that can inform the prioritization, design, and subsequent prototype-based evaluation of standards-compliant Augmented Hybrid TV services. Because the study assessed video-based scenarios rather than functional systems, the results represent perceived usefulness and stated interest rather than observed adoption or technically validated Quality of Experience.**

“This work was supported, in part, by the following grants with refs.: PID2021-126645OB-I00, funded by MICIU/AEI/10.13039/501100011033/ and by “ERDF A way of making Europe”; and CIAICO/2022/025, funded by the Conselleria de Innovación, Universidades, Ciencia y Sociedad Digital of Generalitat Valenciana (DOGV 8919/05.10.2020)”.

F. B., B.R. and S.J.F. authors are with the Universitat Politècnica de València, Spain (e-mails: {fboronat,sflores,broig}@upv.es).

E. S. author is with the Universitat de València, Spain (e-mail: enrique.sigalat@uv.es).

## I. Introduction

THE wide availability and diversity of media content, together with the ubiquity and the multiple connectivity capabilities of modern consumption devices, already enable the combination and coordination of hybrid (broadcast/broadband) delivery technologies for developing and offering new television (TV) services. Although it is technically feasible to provide augmented TV services using only broadcast technologies, all of them require additional bandwidth (scarce and expensive in the broadcast domain), and not all customers may want to pay for (or even -freely-use) them. The use of broadband delivery technologies to provide these augmented services only to those interested customers is a more efficient and cheaper strategy. Although the mentioned heterogeneous media ecosystem may indeed lead to rivalry and incompatibility problems, it also brings fascinating new research and development opportunities.

This paper focuses on the hybrid TV context, in which Internet Protocol (IP) delivery is already having a real impact, offering viewers unrivalled choice and flexibility. Nevertheless, the less noticed and discussed trend is that DVB-based Digital Terrestrial TV (DTT) continues to grow its overall number of TV households over the years and to evolve and perform very well in the market. According to the last *DTT - Global Strategic Business Report* ([1], May 2026), the global market for DTT was estimated at US$4.0 billion in 2025 and is projected to reach US$6.5 billion by 2032, growing at a Compound Annual Growth Rate (CAGR) of 7.1% from 2025 to 2032. According to the latest DTT Industry’s Growth Dynamics and Insights report ([2], February 2026), “*the future growth trajectory of the DTT market hinges on adapting to evolving consumer preferences. Companies need to strategically position themselves by focusing on providing seamless integration with smart TVs and mobile devices, offering value-added services like enhanced interactive features and personalized content recommendations, and exploring hybrid DTT-OTT models to cater to a more fragmented media landscape. The successful navigation of these challenges will determine the long-term market dominance of key players and shape the future of DTT.*”

Despite the shifting dynamics, DTT remains a trusted platform and the primary medium for TV consumption across multiple viewer segments. The evolution to a hybrid TV (i.e., terrestrial, cable, and/or satellite digital TV -hereafter all referred to as DTV- and IP combined) has meant that DTV is now well positioned to underpin the market as it moves towards a pay-lite model. TV consumers can actively supplement linear DTV with broadband IP-based services, thereby accessing catch-up media and subscription video-on-demand (SVOD) platforms such as Movistar, Vodafone, Netflix or Amazon. The success and viewers' appetite for the hybrid model suggest that an IP-only future is still a long way off. Instead, continuing to blend the best of IP and DTV into a workable hybrid offer that meets today's consumer needs remains a relevant area of technical and user-centered research.

For audiences, the result is a very reliable service built around a core (DTV) broadcast service, with an enriched set of IP-delivered services. This offers flexible ways to watch TV (even across multiple devices), including a more accessible, interactive and increasingly personalized and engaging viewing experience. Broadband (i.e., IP-based) delivery can provide tailored content and ensure customization and accessibility in immersive TV viewing experiences. Accessibility services can be easily offered in this model to address the needs of those, e.g., with hearing and low-vision problems or learning difficulties, as well as older people. For example, adaptive subtitles in different languages, audio descriptions, and/or sign language videos in different formats can be provided, via broadband, as complementary content to the main (broadcast) TV content. Moreover, users can personalize the playout of this kind of content (e.g., larger subtitles, placed near the speaking character, or audio descriptions presented as text). Some examples of basic hybrid TV services are the following (explained in [3]): provision of additional audio streams via broadband technologies; multi-view (or *Free ViewPoint*) TV; remote Participation in TV shows; Social TV (a.k.a. *Watching Apart Together*); provision of spatial, temporal and color scalability via broadband to augment the quality of the broadcast TV content; provision of advanced functionalities (e.g., catch-up, re-plays, fast rewind/forward…); targeted and personalized advertisements; or integration of web feeds (e.g., quizzes, relevant information or statistics).

Some standardization and research initiatives to drive these services, such as Hybrid Broadcast Broadband TV (HbbTV [4]), Advanced TV Systems Committee (ATSC [5]) and MPEG Media Transport (MMT [6]) standards, the former Smart TV Alliance [7] and some EU-financed projects (mainly based on the HbbTV standard, such as HBB-NEXT or HBB4ALL), focused on using hybrid technologies to offer enhanced TV services. Thanks to those efforts, a new wave of enriched services and even new business models is already possible. This is particularly relevant to TV operators and other stakeholders. In [3], the *hybrid TV media services* concept was introduced, which refers to interactive media services that concurrently use broadcast and broadband technologies to consume TV-related media content, either in synchronization with or independently of the main broadcast TV content. This content can be consumed on a single device (a.k.a. hybrid terminal, e.g., a connected TV) or on multiple devices in multi-screen scenarios (e.g., one or more hybrid terminals and/or one or more companion devices). Nevertheless, the currently implemented solutions still do not fully exploit the potential and opportunities that hybrid TV services can offer.

Until now, DTV content has been viewed in 2D or 3D, offering scenes and points of view selected by the production team. However, the introduction of more modern types of content, such as omnidirectional or 360º video, CGI (Computer-Generated Imagery) and Virtual or Augmented Reality (VR or AR) content, in the TV services will allow the TV service providers to offer more immersive viewing experiences (hereafter referred to as *Augmented Hybrid TV services*). Omnidirectional (i.e., 360º) and AR/VR content will not replace traditional TV content, but can complement it, providing more immersion. If 360º video content is available, users can choose their point of view or field of view (FoV) for each scene. All this is already possible thanks to the proliferation of consumer devices (smart TVs, tablets, and AR/VR glasses or Head-Mounted Displays -HMDs-) and their ability to connect to heterogeneous access networks. Evidence of the impact of AR and VR technologies on current and future society includes findings from recent studies and recent moves by major technology companies in this field. According to a recent report in [8] (March 2026), the global AR and VR Market size was valued at approximately USD 22 billion in 2024 and is expected to reach more than USD 268.58 billion by 2032, growing at a CAGR of 18.4% during the forecast period 2026-2032. Facebook (Meta), Microsoft, Google, Apple, Sony, HTC, and Samsung have bet heavily on VR/AR technology, reflecting its high potential. AR/VR technology will not focus solely on video games, but has significant potential in other areas, such as education, health, and TV. In recent years, the Spanish public broadcaster RTVE has already experimented with VR and interactivity in TV series '*The Ministry of Time*' and '*The Case*'.

On the one hand, Figure 1 illustrates the main concept of an immersive hybrid TV media system that supports Augmented Hybrid TV services. While the main TV content is broadcast using DVB technology, complementary content can be available on remote media servers accessible via the Internet. According to the HbbTV standard terminology, broadcast content is consumed in the so-called *Main Screen* (MS) app, and complementary content is consumed in the so-called *Companion Screen* (CS) app. On the other hand, Figure 2 shows an example of a domestic multi-device, immersive hybrid content consumption scenario (i.e., the end-consumer part of the system shown in Fig. 1), including 360º video, AR, Computer-Generated Imagery (CGI), and 3D content (hereafter, the term *immersive content* is used to refer to all of them). The TV set receives, via broadcast (e.g., DTV or IPTV), a video stream of a part of the scene (selected by the production team), and additional content is received via broadband networks. In the figure, the latter consists of 2 video streams of 2 areas of the scene (selected by users -*personalization*-) displayed on a tablet and a laptop, and a 360º video stream of the complete scene to be viewed on an HMD or in a 360º video player on a flat 2D screen. The user can watch the entire scene by turning his head or dragging his finger/mouse across the screen, respectively, and then selecting a new FoV. CGI can provide additional information about omnidirectional content (e.g., arrows indicating the scene's main FoV at each moment). The user can also privately view TV-related content on virtual screens using an AR HMD (e.g., 3D models, personalized subtitles, sign language videos, or complementary views), as shown later. To further enhance immersion and the user's QoE, sensory effects (e.g., smell, vibration, wind, or cold/heat generators) can also be included in that scenario to stimulate senses other than sight and hearing.

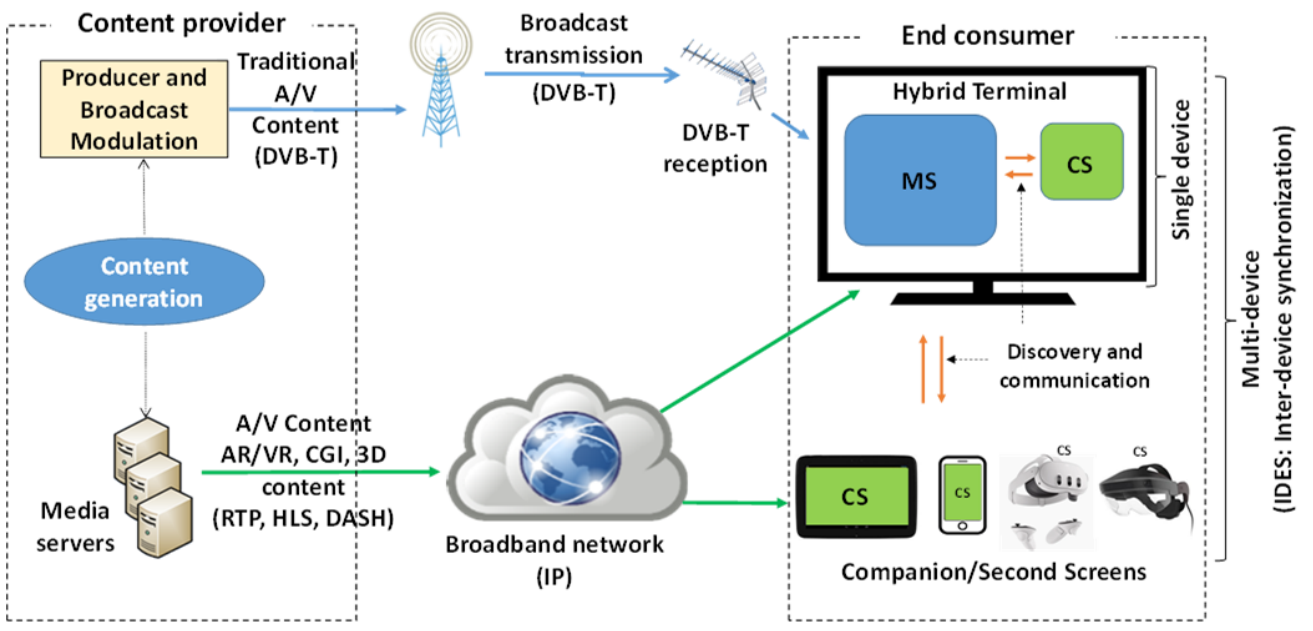

Figure 1. Overall Immersive Hybrid TV system

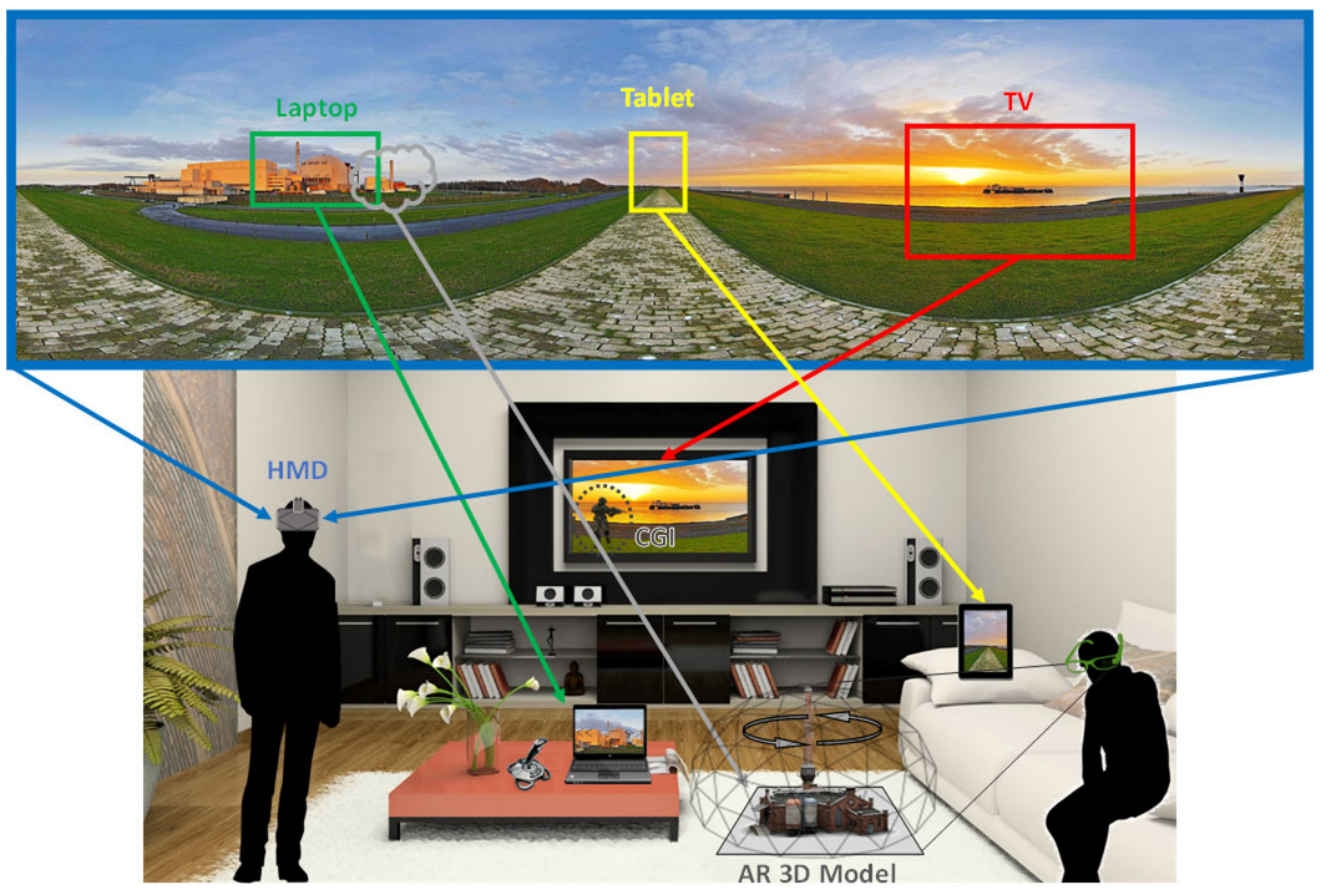

Figure 2. Multi-device hybrid TV consumption scenario, including immersive/accessible content

To guarantee an acceptable user's QoE in Augmented Hybrid TV services, accurate synchronization of the playout of MS and CS apps is needed. The delay variability in hybrid TV services and the need for synchronization solutions (a.k.a. *hybrid sync* solutions) were discussed in the authors' previous work in [3]. The hybrid sync term is used when multiple related streams, delivered via hybrid broadcast/broadband technologies, need to be played out synchronously. It can involve one or more well-known media synchronization types to ensure in-sync playout of these related streams [9]. The synchronization of the playout processes of related streams is known as *inter-stream synchronization* (e.g., lip-sync, i.e., the synchronization of audio and video playout in a media presentation). The streams involved must be played synchronously, either on the same device (a hybrid terminal if the streams are received via different delivery technologies, such as a Smart TV) or across different devices. When streams are played on different devices co-located in a local environment (e.g., the same room, in a multi-screen scenario), this type of synchronization is known as *inter-device synchronization* (IDES, see Figure 1). Timing artefacts introduced by agents in heterogeneous end-to-end multimedia distribution chains must be corrected across all related devices to ensure in-sync presentations and good QoE.

This new enriched TV media consumption paradigm and immersive hybrid TV media services, on the one hand, can have a significant impact, e.g., on entertainment, educational programs or TV marketing. On the other hand, they can also bring social benefits, enabling users to feel more integrated (especially those people with AV impairments) and immersed while consuming media.

Currently, we are witnessing a rise in the use of multiple devices (e.g., to connect to social networks and/or browse) while watching TV. According to the recent Connected TV Study in [10] (May 2026), conducted by IAB Spain, on the one hand, 95% of Spanish Internet users aged 16-75 use AV content services via the Internet for TV (i.e. connected TV), regardless of the access device, representing around 34.3 million individuals. 53% of them use other devices while watching TV (mainly social networking -76%-, Internet browsing -55%-, or looking for information about the advertising they watch on Connected TV -9%-). Series and films are the leading content on Connected TV. Traditional content (news, game shows) is losing ground to on-demand entertainment; however, there are differences by gender and generation: sports are consumed more by men and older adults. Regarding advertising on connected TV, only 5% admit to having scanned a TV-advertising QR code. They use those codes as a complement, not as their primary means of interaction. On the other hand, people from professional sectors participating in that study admitted they will continue to increase their offer of Connected TV services for two main reasons: it is a way to reach audiences that are increasingly consuming less traditional media (65%), and to adapt to the current AV consumption paradigm (59%). 90% of professionals expressed interest in the competitive advantages of Connected TV as an advertising medium. Connected TV is primarily used to expand coverage and improve segmentation, while more advanced uses, such as retargeting or cross-device strategies, are still secondary in Spain.

Despite the above studies, to the best of the authors' knowledge, there is no scientific knowledge base on the receptivity, habits, requirements, preferences, and needs of both consumer and professional sectors regarding *Augmented Hybrid TV services* in any country. New studies are still required to determine them, as well as the requirements/characteristics of the most appropriate immersive contents to be included and how they should be included (combination) in the prototypes to be developed in the future. Moreover, there is no consensus on the signaling mechanisms for immersive content (notification of availability, characteristics, dependencies with other content...), nor on their possible interaction/presentation modes. Likewise, their inclusion can affect production processes and classic AV narratives (e.g., cuts between scenes and 360º video). It is necessary to research this area to provide recommendations for adequate preparation, integration, interaction, and presentation of immersive content. For example, friendly notifications (e.g., specific CGI alerts or on-screen cues) could be required to inform users about the availability of immersive content.

The authors' previous study in [3] provided statistical data and some interesting insights into 4 basic use cases of hybrid TV services, their relevance, and the need for better technological solutions to support them (especially regarding IDES). The selected use cases were: 1) Online sports commentator: a consumer watching a sports match on TV, while using a companion device to listen to an alternative commentator from an online radio channel; 2) Intelligence quotient (IQ) test TV program; 3) Multi-view camera in Formula 1/Moto GP (Grand Prix); and 4) Social TV. More than 1,000 Spanish consumers participated in the study, who were asked about their viewing habits, needs, preferences, and expectations regarding these types of services. More than 90% of them reported experiencing delays (i.e., lack of synchronization) in these scenarios, which they considered a serious barrier. Several challenges must still be addressed to deliver hybrid multimedia content in a timely manner so it can be played simultaneously on all devices involved. The above use cases are a reality nowadays, and traditional devices (tablets or mobile phones) can serve as secondary devices for displaying complementary content.

Nevertheless, to the best of the authors' knowledge, no AR/VR-based hybrid TV services are commercially offered yet. AR/VR technologies are becoming increasingly commonplace, and the cost of AR/VR devices is decreasing. This will help to boost immersive TV consumption experiences. Nevertheless, even though hybrid systems and immersive media have been extensively studied from a technical perspective, the analysis of user acceptance of augmented hybrid TV systems (i.e., their effects on users' QoE) remains underexplored, and assessment using functional prototypes is needed.

In this paper, the continuation of the authors' previous study in [3] is presented but now focusing on the next evolutionary step in the TV hybrid ecosystem, where immersive media is integrated with the traditional TV experience and multi-device environments. The Spanish consumers' perspective on five relevant use cases of Augmented Hybrid TV services (described in Section IV) has been addressed. An exploratory stakeholder consultation and co-design workshops, along with a survey, have been conducted to obtain valuable insights into receptivity, habits, preferences, and needs from consumers (gender-balanced and including people with AV impairments) and from a small group of AV professionals, regarding the inclusion of immersive content in the Augmented Hybrid TV viewing experience.

Three workshops conducted at the beginning helped to select several potential use cases for Augmented Hybrid TV services to be considered in the study. One workshop was conducted with professionals from the TV media ecosystem (regional and TV operators), while the other two were conducted with people from national and local institutions that provide services to people with AV impairments. The stakeholder workshops informed the selection and formulation of five representative use cases (described later). Then, a survey was designed and conducted via an online questionnaire that included questions on key aspects of those use cases and other segmentation variables (general sociodemographic characteristics, specific variables related to TV content consumption, etc.). Each use case was evaluated across different segments of potential Spanish users.

The study is conceived as an exploratory user requirements and preliminary acceptance assessment of those representative use cases for Augmented Hybrid TV services. Rather than evaluating implemented systems or observed market adoption, it examines perceived usefulness, stated interest, anticipated benefits and limitations, and user-centered priorities for subsequent system design and prototype-based evaluation. The results reveal valuable insights, including high consumer interest in the selected use cases and their identified benefits, despite their current lack of availability and identified limitations. Professionals participating in the exploratory sessions expressed concern that production and maintenance costs could limit adoption by smaller broadcasters, and that further research is still needed to overcome key remaining challenges, with a special focus on technical challenges (e.g., hybrid sync or bandwidth scarcity). They also emphasized that some form of training will be needed to accelerate the adoption of Augmented Hybrid TV services, especially among users lacking the digital competencies needed to configure and use secondary devices and second-screen applications.

Although this study was conducted among Spanish consumers and professionals, Spain is one of the main European countries where hybrid TV media services are regularly available and are being progressively adopted and improved. The results and conclusions can be useful not only to Spanish stakeholders but also to other international stakeholders where similar initiatives to drive Augmented Hybrid TV media services are underway. The authors expect the contributions of this paper to be valuable to agents in the global media consumption landscape as they envision what remains to be developed, offered, and improved in next-generation hybrid TV media services.

The main contributions of this work can be summarized as follows:

- Definition of Augmented Hybrid TV services. Five representative use cases that combine traditional TV with AR/VR technologies and multi-device interaction are described and analyzed.
- Large-scale evaluation of potential Augmented Hybrid TV services. The work presents the results of an exploratory survey of more than 600 Spanish people. It is the most comprehensive analysis of Spanish users' opinions and expectations for these kinds of services.
- Comparative evaluation of the selected use cases. They are

evaluated, enabling a systematic comparison of user perceptions across different categories of Augmented Hybrid TV services.

- Identification of key perceived facilitators and anticipated barriers for Augmented Hybrid TV services. Accessibility features, usability, technological familiarity, and device ergonomics have been identified as key factors influencing user acceptance.
- Translation of user-perceived requirements into indicative priorities for hybrid TV system design. Building on the preceding findings, the study derives a structured correspondence between the empirical evidence gathered, the design priorities that participants implicitly or explicitly identified, and their indicative implications for hybrid TV system design. This correspondence is intended as a user-perspective reference to inform, but not substitute for, subsequent technical design and standardization decisions.

The structure of the paper is as follows. In Section II, the concept of Augmented Reality TV is presented. In Section III, the conducted exploratory stakeholder consultation and co-design workshops are described, and the insights obtained from them are briefly explained. In Section IV, the five use cases of Augmented Hybrid TV services considered in the study are described. Then, in Section V, the research survey is presented. The study design, methodology, instruments, participant sample, and general data are briefly described. Section VI summarizes the results obtained, and a brief discussion of them and some of the study's limitations are presented in Sections VII and VIII, respectively. Finally, the paper concludes in Section IX.

## II. Augmented Reality TV

The convergence of TV and immersive technologies has given rise to a field broadly known as *Augmented Reality TV* (ARTV), which encompasses the integration of TV with AR, VR, and other computer-generated content [11] [12]. ARTV is defined as '*an immersive, real-time interactive digital experience in 3D centered on the TV set*' [13]. In [12], the first conceptualization of ARTV was presented, and several possibilities for ARTV systems and applications were identified, such as conventional world/virtual TV or augmented world/conventional TV. Connections between ARTV, as a distinct area of research, and 3D TV, ambient media, and multimedia alternate realities were also described. In [14], an agenda for researchers to consider when designing for ARTV was proposed. ARTV can manifest in various ways, from complementary information displayed on mobile devices to spatial projections that augment the viewing environment or fully immersive experiences using HMDs [15]. Preliminary investigations into users' attitudes indicated a strong interest in these enriched and immersive experiences [13] [16]. Early efforts focused on improving the viewing experience by extending visual content beyond the TV screen. In [17], a predictive approach was introduced to continue the primary video stream across neighboring displays, thereby delivering a wider and more immersive cinematic environment. Although these projection-based solutions proved useful for generating an ambient peripheral effect, they suffer from several important drawbacks. Specifically, they require complex calibration procedures tailored to the characteristics of each physical space and, more importantly, they provide the same experience to all viewers present, without supporting individualized personalization.

The increasing demand for more personalized and engaging experiences in consuming TV-related content has spurred the emergence of second-screen applications running on personal devices. In [18], a hybrid system was proposed in which metadata synchronized with the broadcast stream enables the rendering of 3D objects on the TV, while a companion device provides interactive control. Similarly, in [19], another system was introduced in which animated 3D elements appear to emerge from the TV display and are visualized through a smartphone. Although both works demonstrated the viability of the hybrid approach and highlighted its capabilities for interactive experiences, no user-based evaluations were conducted. Nevertheless, those experiences remain largely confined to conventional 2D handheld screens and fail to fully leverage the spatial and immersive features of next-generation AR/VR devices. More recent research has begun to leverage HMDs to create truly immersive AR companion experiences. In [20], the *MR TV Mozaik* system was introduced, enabling users to watch TV through an AR HMD while interactive elements and controls are arranged as a virtual mosaic surrounding the physical TV display. This work demonstrates the potential to integrate 3D objects and contextual information into the user's surroundings, thereby enabling richer forms of interaction. The *MR TV Mozaik* framework was further expanded to support collaborative experiences in [21]. However, this work also lacked a user-centered evaluation. In [13], the preferences of 172 European participants (mostly from Romania, but no data about their distribution were provided) regarding ARTV were analyzed, and results from an exploratory investigation examining the perceived usefulness of 20 different ARTV scenarios were presented (many of them based on projections on the walls surrounding the TV device). The participants were not exposed to an actual ARTV service; instead, they were asked to complete an online survey after viewing four brief illustrative YouTube videos about ARTV. They perceived ARTV as more useful, fun, and desirable than discomforting or cognitively demanding. Participants rated the 20 use cases by using a 5-point Likert scale with the following items: '1, not valuable', '2, limited value', '3, moderately valuable', '4, valuable', and '5, very valuable'. On the one hand, high value was perceived for ARTV scenarios involving interactive content, wall- and room-sized video projections, multiple perspectives of the same movie scene, and virtual objects emerging from the TV screen into the room. On the other hand, scenarios such as movie subtitles shown outside the TV set, live video or 3D representations of remote friends watching the same broadcast, or multiple virtual channels displayed around the TV screen were rated less valuable. Subsequently, in [22], the same researchers conducted a similar survey involving 147 Chinese participants to examine potential cultural differences in viewers' expectations regarding ARTV. They compared the results with the preferences expressed by participants from Europe in the original study [13]. Application scenarios that combine AR and TV were perceived as valuable across cultures (mean ratings > 3), but rankings differed within each culture.

In [16], an experimental study involving 12 participants explored the augmentation of TV content with synchronized AR holographic elements. In the developed prototype, user interaction with the AR content was limited. The findings indicated that enriching existing TV programs with AR components can enhance viewer engagement, while the prototype itself was regarded as innovative, stimulating, visually attractive, and easy to understand. These findings further motivate the investigation of robust, scalable, and standards-compliant mechanisms for delivering and synchronizing complementary immersive content with conventional and hybrid TV services.

In [23], the potential of AR technologies to enhance TV viewing by synchronously delivering sign language interpretation was investigated. An exploratory study involving 23 participants (11 from the UK and 12 from Germany) examined three approaches to watching a sign language interpreted program. These comprised a conventional in-vision technique (i.e., PiP), where the interpretation was displayed directly on the TV screen, and two AR-based alternatives, in which a virtual sign language interpreter (a 'half-body' representation and a 'full-body' representation) was positioned just beyond the TV screen's boundaries. The study also analyzed participants' preferences and explored the differences observed between both countries.

In [15], the user experience (UX) of 14 participants watching TV while transitioning between four levels of augmentation was investigated. The study analyzed four ARTV configurations: a virtual TV screen; AR content displayed outside the TV frame; AR elements floating in the space between the TV screen and the user; and an augmented living room with immersive TV. The findings indicated a UX characterized by high perceived usability, low-to-moderate cognitive workload, strong levels of engagement and comprehension, and a high degree of perceived involvement with the content, together with a moderate sense of transportation and detachment from the physical environment.

One of the primary technical challenges in any multi-screen scenario, and particularly in ARTV scenarios, is preserving temporal synchronization between the playout processes of the main and companion devices, when required [24]. A lack of synchronization between these processes can significantly impair users' QoE [25]. Previous work has introduced several approaches to address this issue, including web-based architectures designed for event-driven communication [26], synchronization strategies based on user feedback [27], and specialized metadata designs for coordinating broadcast and AR elements [28].

In summary, previous work has explored a range of implementation strategies for ARTV services, from projection-based systems to companion-screen solutions, while also examining users' preferences for ARTV scenarios, reporting engagement levels with ARTV, and presenting initial basic user ratings for specific use cases. They do not consider accessibility issues or 360º video in ARTV services, and only one focused on HbbTV-related services (the part of the German study, with only 12 participants, in [23]). In this paper, a larger user-centered exploratory study is presented following an approach based on five potential ARTV use cases in which linear broadcast TV is augmented with broadband media content that can be consumed in AR/VR-based second-screen applications. The use cases have been carefully selected, and insights have been obtained for each. This study also aligns with the first item of the ARTV research agenda proposed in [14], '*Understanding users' preferences*'.

## III. Exploratory Stakeholder Consultation and Co-Design Workshops

The following three workshops were conducted to consult potential stakeholders and select the use cases of Augmented Hybrid TV services to be included in the study:

- Workshop with 7 professionals working in the AV media sector, who provided perspectives related to content production, media distribution, and audience engagement.
- Workshop with 6 people from the ONCE[1], the *Spanish National Organization for the Blind*. It is a Spanish foundation that supports people with blindness or severe visual impairments in Spain and is internationally recognized for its work in social inclusion.
- Workshop with 8 people from APESORGA[2], the *Association of Deaf People of Gandia*. This non-profit organization promotes accessibility and inclusion for individuals with hearing impairments and their families in the La Safor region of Valencia, Spain, where the authors' Institution is located.

Each workshop was organized in two approximately 40-minute sessions. The first session introduced the concept of Augmented Hybrid TV services and included a brainstorming activity to identify potentially relevant use cases. In the second session, the research team presented the most representative and feasible scenarios for structured discussion and assessment. Their most relevant benefits and limitations were identified. The last two workshops offered valuable additional perspectives from individuals with AV impairments, enabling the authors to better address accessibility requirements in the selected survey use cases.

The discussions across the workshops revealed several converging perspectives on the usefulness and the feasibility of some potential Augmented Hybrid TV services. The possibility of providing multilingual adaptive and personalized subtitles via broadband for main TV content was consistently ranked among the most useful services by both AV professionals and APESORGA participants. All participants emphasized the significant accessibility benefits of customizable subtitles, particularly in facilitating language learning, improving comprehension in noisy environments, and enabling more private reading experiences. However, participants also highlighted that precise synchronization

[1] In Spanish: Organización Nacional de Ciegos Españoles. https://www.once.es/ (last access: September 2026)

[2] Association of Deaf People of Gandia "La Safor" (In Spanish: Asociación de Personas Sordas de Gandia "La Safor", APESORGA). http://www.apesorga.org (last access: September 2026)

between subtitles and AV content is essential, as delays could undermine narrative coherence.

On the one hand, the use of an AR HMD to display complementary video content (e.g., sign language interpretation or contextual statistics) was generally ranked the second most useful feature in the three workshops. The participants appreciated the ability to access additional information without interfering with the main viewing experience or affecting other viewers in the same room. This approach was seen as particularly valuable for accessibility services and contextual information in sports or educational content. Nevertheless, several potential drawbacks were identified, including the risk of information overload, attention fragmentation, and the relatively high cost of AR devices. On the other hand, the use of a VR HMD to display complementary 360° video content was rated as moderately useful, particularly for entertainment and sports applications. Participants appreciated the opportunity to explore immersive environments and choose their own viewing perspective. However, both APESORGA participants and AV professionals reported potential adverse effects, such as dizziness or motion sickness, when using VR headsets. In addition, the technology was perceived as potentially isolating viewers from shared TV consumption experiences.

The possibility of projecting a large virtual TV screen (as a substitute for the actual TV screen in the room) using AR received particularly strong support from the ONCE participants. They ranked it as their most useful use case. Participants highlighted the potential benefits for people with low vision, since the virtual screen could be enlarged and adapted to individual visual needs. AV professionals ranked this use case as moderately useful, whereas participants from APESORGA considered it less relevant for their needs. Concerns raised during the discussions included potential visual fatigue from prolonged use and the high bandwidth required to maintain good video quality on high-resolution, immersive, or virtual displays.

An additional augmented TV service raised in the workshops was the possibility of providing interactive 3D objects in AR/VR to complement the main broadcast TV content. Although participants acknowledged its potential for advertising, educational applications, and immersive demonstrations (e.g., exploring the interior of a vehicle), concerns were raised about the risk of distraction from the main TV content and the relatively high production costs of creating interactive 3D assets. Both the ONCE group and the AV professionals ranked it as the lowest-priority scenario.

Despite differences in priorities among participants in each workshop, several common themes emerged during the discussions. First, synchronization between broadcast content and augmented virtual elements was repeatedly identified as a critical requirement for maintaining narrative coherence and usability. Second, accessibility and personalization were highlighted as key drivers of value for Augmented Hybrid TV services, particularly among users with visual or hearing impairments (i.e., from APESORGA and ONCE). Finally, participants recognized the strong potential of immersive technologies to expand TV storytelling and enable new forms of interactive content. However, concerns remain regarding device comfort, visual fatigue, and the cost of devices or hardware.

All the participants in the workshops agreed that integrating sensory effects into TV consumption experiences would enhance immersion and engagement, thereby improving the quality of their experiences. It was emphasized that sensory effects could help people with disabilities or functional diversity (*social integration*). Some examples are: scents, wind, and temperature can be used to inform people with hearing or visual impairments about the scene without requiring a description (e.g., text or audio-based one); in a film, when a character opens a can of orange soda, after the opening sound the orange smell is present, helping them to identify the can content; when in a TV scene someone opens a window and a fan generates wind effect; or when someone lights a bonfire and a sensation of heat is felt.

Professionals noted difficulties in providing multisensory TV experiences to consumers due to a lack of standardization, the cost of producing sensory metadata, and the cost of compatible effect-generation devices for home use. Nevertheless, they all agreed that the MPEG-V standard and the proliferation of IoT-based devices would help boost such experiences.

Overall, the results from the workshops confirmed the strong perceived value of accessibility-oriented services and the conditional acceptance of immersive technologies, contingent on usability and technological maturity.

## IV. USE CASES OF HYBRID TV SERVICES SELECTED FOR THE STUDY

After a thorough discussion and based on insights from the workshops, the research team selected five representative use cases of Augmented Hybrid TV services for the survey. They were selected because they represent different levels of immersion and technological complexity within augmented TV ecosystems and are listed and described below.

1) *Adaptive and personalized subtitles, which can be displayed either on the TV screen or on secondary devices (tablet, mobile or AR HMD).* Users can watch customized subtitles (obtained via broadband) in one or more languages, either on the main TV (hybrid terminal) or on secondary or companion devices, synchronized with the broadcast content watched on the main TV (Figure 3). Subtitle features can be customized, such as their position on the screen (or out of the screen if an AR HMD is used), text format (font, size, color, etc.), background color (for greater contrast), in one or several languages (i.e., at once), in one or multiple lines, with or without additional delay, etc. (Figure 3a). This can be useful, for example, for people with AV impairments or for learning foreign languages, among other purposes. Figure 3b shows an example of an AR HMD in use. In that case, the subtitles would only be visible to the user wearing it and would not be intrusive or annoying to other users watching TV in the room. In that case, the user could position the subtitles off-screen, so they would not overlap the main TV content.

2) *Complementary (broadband) video content displayed on an AR HMD, providing additional information synchronized with*

*the main (broadcast) TV program.* Users can consume broadcast 2D AV media content on the main TV and complementary content (e.g., customized subtitles, sign language video, other views, statistics or audio guides), received via broadband, on secondary or companion devices (e.g., tablet, smartphone or AR HMD). An example of an AR HMD in use is shown in Fig. 4 (overlaps with the TV main content can be avoided).

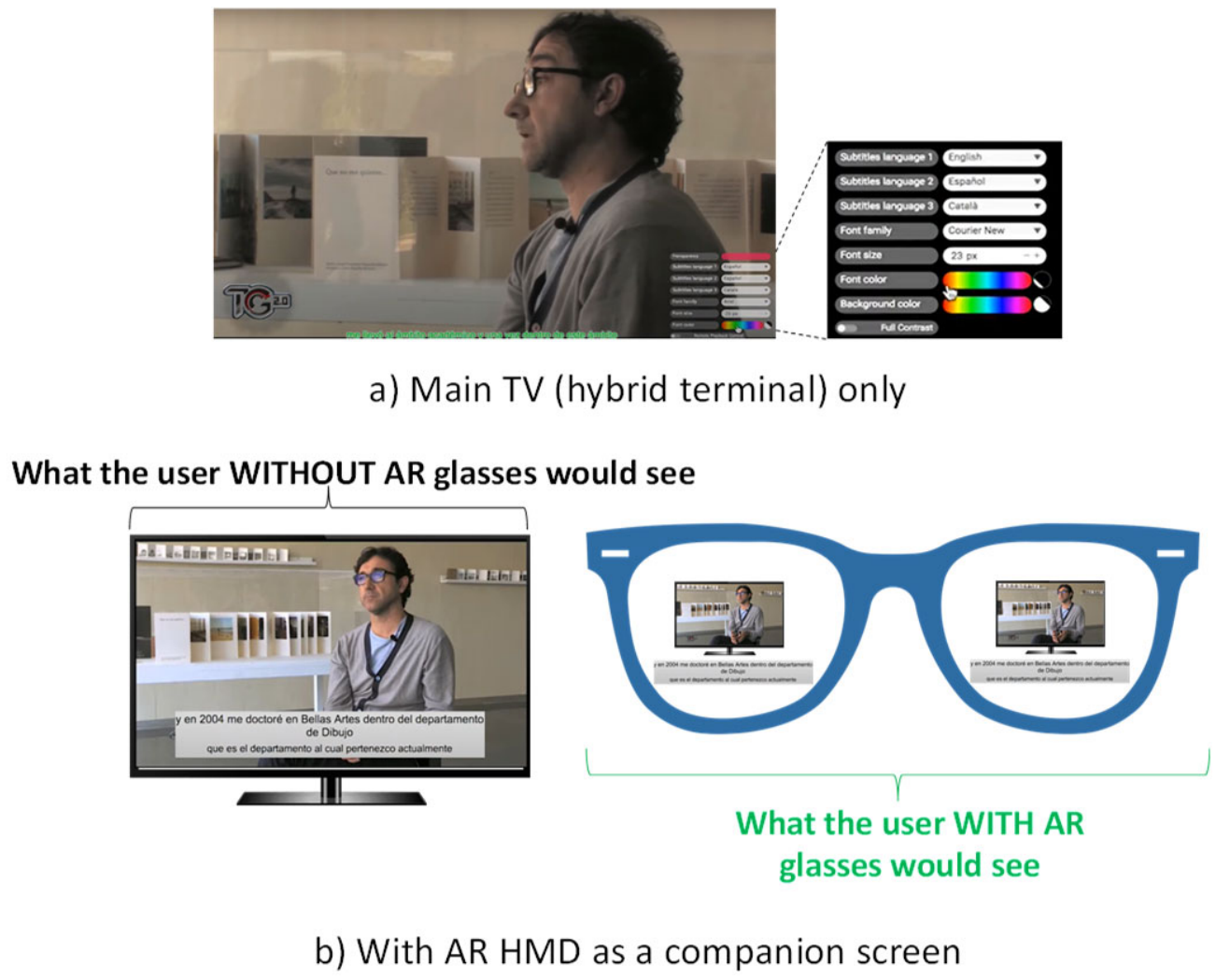

Figure 3. Customized subtitles received via broadband

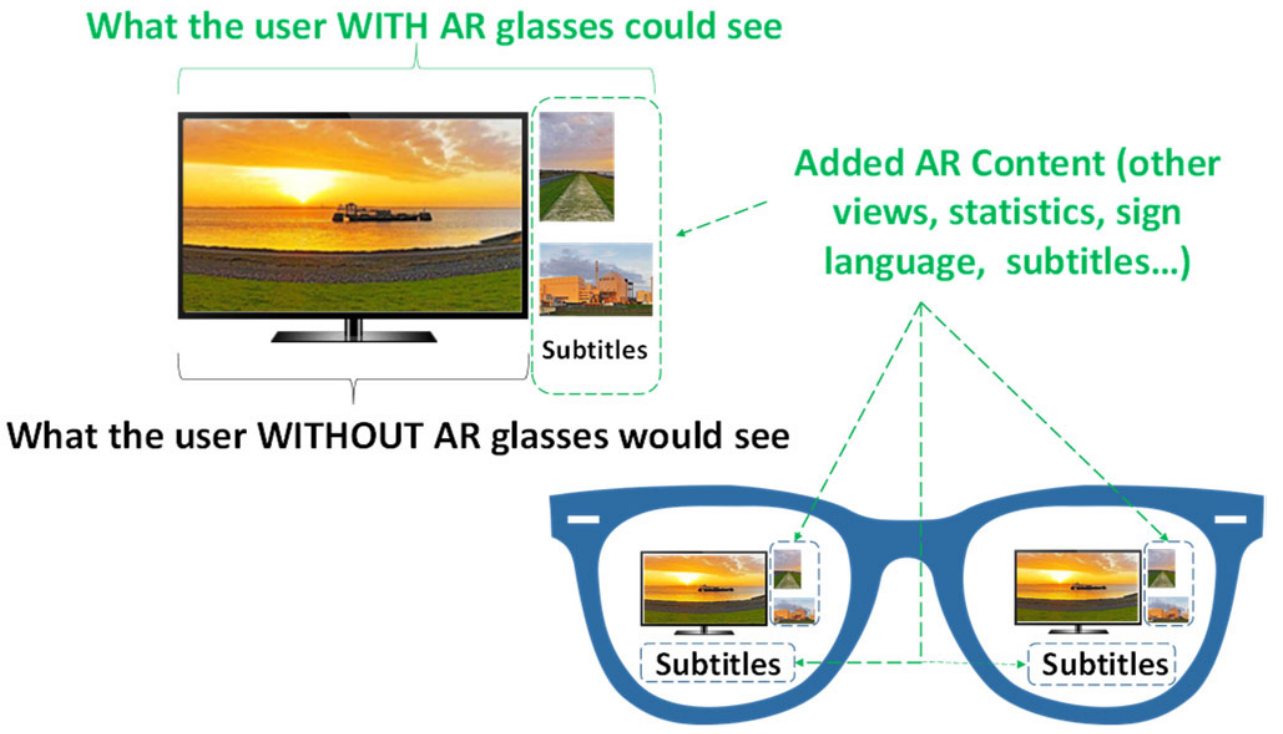

Figure 4. Complementary and accessible content in an AR HMD

3) *Virtual large-screen TV viewing through an AR HMD, allowing users to experience TV content on a virtual screen larger than the physical TV in the room.* Users can watch broadcast 2D AV media content on the main TV and/or the same content, but received via broadband, on a larger virtual screen by using an AR HMD (Figure 5). This case can be useful for family members with any visual impairment who are watching TV in the same room as other family members and talking about the TV content (i.e., interacting) with them.

4) *Combination of traditional (broadcast) 2D TV content with synchronized (broadband) virtual 3D objects/models/animations presented in secondary devices, enabling spatial visualization of complementary information.* Users can watch broadcast 2D AV media content on the main TV, and related 3D virtual objects, 3D models, or VR animations on companion screens (e.g., tablet, smartphone, or AR/VR HMD). They can interact with 3D models or VR animations of, for example, products advertised on TV (e.g., zoom in/out, rotate, move, etc.). Figure 6 shows an example of using different types of companion screens to view complementary 3D models or animations of a car being advertised on TV.

5) *Combination of traditional (broadcast) TV content with complementary (broadband) 360° video, allowing viewers to explore immersive scenes related to the broadcast TV content.* Users can watch broadcast 2D AV media content on the main TV and a complementary 360º video on a VR HMD (Figure 7). For example, imagine a father and his son watching a football match on TV together in the same room, but the latter is watching the match in a 360º video (e.g., from the point of view of someone sitting in a stadium seat), and both are talking about the match.

In all the use cases, to provide good QoE for users, the playout processes for the content being watched on the TV and on the companion screens must be accurately synchronized.

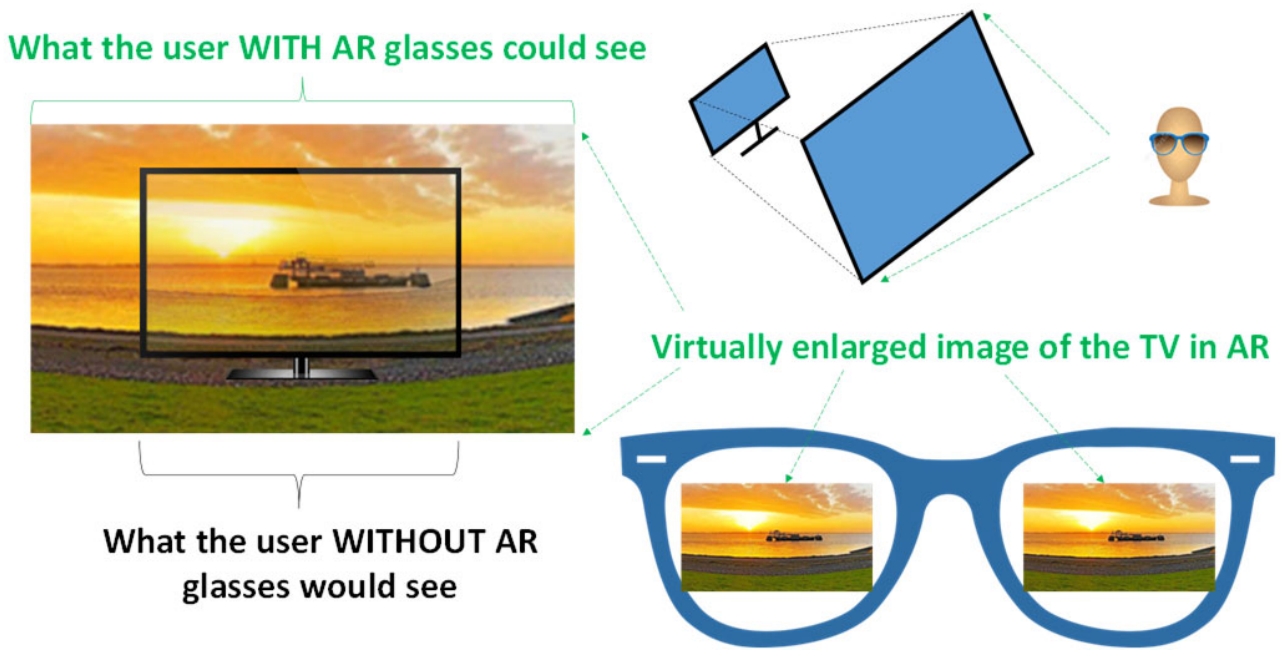

Figure 5. Bigger virtual TV screen in AR HMD

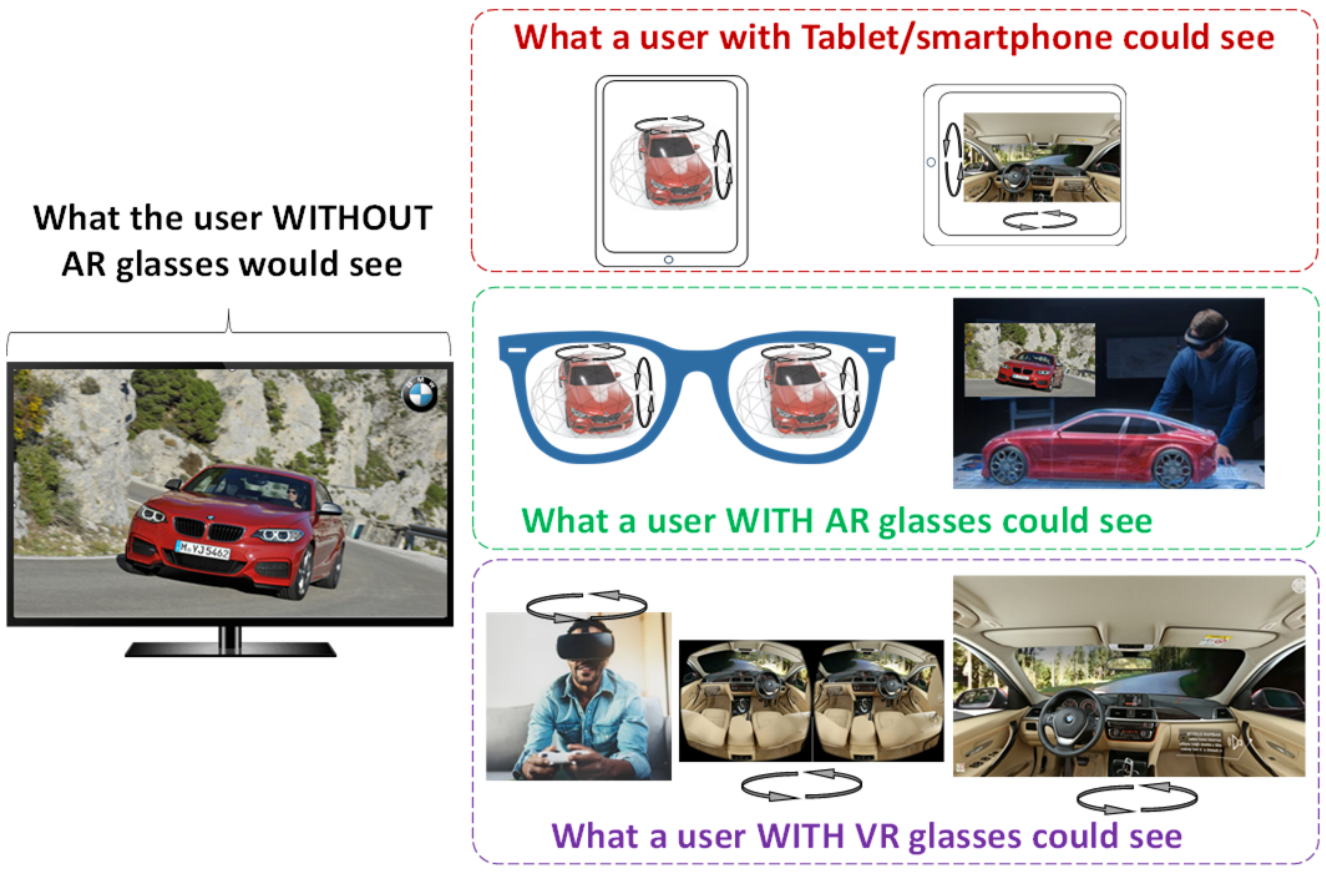

Figure 6. Complementary 3D models and animations

## V. Empirical Survey Methodology

This section describes the research design, survey instrument, sampling strategy, quality-control procedures, statistical analyses, and characteristics of the achieved sample.

## A. Research Design and Survey Procedure

A quantitative, descriptive, and cross-sectional research design was adopted to conduct the exploratory user requirements and preliminary acceptance assessment of the five selected Augmented Hybrid TV use cases. Data were collected through a Computer-Assisted Web Interviewing (CAWI) survey administered through a professionally managed opt-in online panel. Fieldwork was conducted in Spain from November 1 to December 31, 2025, in collaboration with Ikerfel S.A.[3], an independent market research agency responsible for panel management, participant recruitment, quota implementation, and fieldwork quality control.

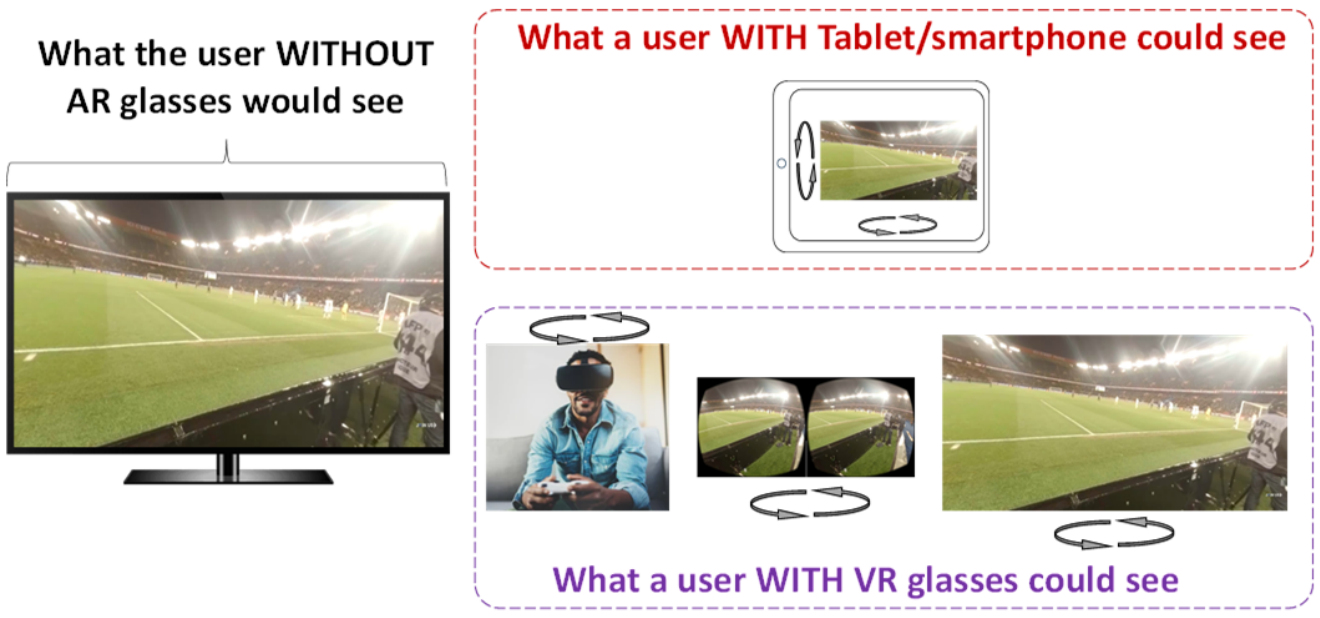

Figure 7. Complementary 360º video

The authors designed a structured online questionnaire[4]. At the beginning of the survey, participants received a standardized video-based explanation of the concept and main technical features of Augmented Hybrid TV services. After reviewing, accepting the privacy and personal data processing policies[5] and providing informed consent, each participant viewed illustrative video demonstrations of two of the five selected use cases described in Section IV. The two use cases and their presentation order were randomly assigned within the questionnaire to reduce respondent burden and potential order effects. Consequently, each use case was evaluated by a partially overlapping subsample of 234-293 participants, as shown in Table 1.

## B. Survey Instrument, Quality Control, and Statistical Analysis

Because the evaluated services are not yet commercially available as fully integrated hybrid broadcast/broadband AR/VR systems, respondents assessed standardized video demonstrations and multimedia descriptions rather than functional prototypes. The questionnaire covered five areas: 1) sociodemographic characteristics; 2) technological familiarity, digital literacy, and previous AR/VR experience; 3) second-screen and multi-device viewing habits; 4) preferences regarding multisensory effects; and 5) perceived usefulness, benefits, limitations, and stated interest associated with the assigned use cases. Perceived-usefulness items were generally measured using five-point Likert-type response scales, whereas other response formats were used where appropriate.

The questionnaire included embedded attention checks and response-quality controls to identify incomplete, mechanically patterned, inattentive, or implausibly rapid responses. The mean completion time among the retained responses was 27 min 55 s, with observed completion times ranging from 23 min 32 s to 35 min 21 s. Statistical analyses were conducted using IBM SPSS Statistics[6]. Descriptive analyses included frequencies, percentages, means, standard deviations, and cross-tabulations. Associations between gender and mutually exclusive categorical variables were examined using Pearson's chi-square ($\chi^2$) tests. When appropriate, statistically significant omnibus associations were followed by two-tailed pairwise comparisons of column proportions using z-tests. For multiple-response questions, each option was treated as a separate binary outcome, and gender differences in selection proportions were examined individually. A Bonferroni adjustment was applied across all categories or response options within the same conceptually related family.

Gender differences in the mean scores of the Likert-type items were examined using two-tailed Welch's independent-samples *t*-tests. Welch's test was selected because it does not require equal population variances and is robust to unequal group sizes. The nominal significance level was set at $\alpha = 0.05$. To control the family-wise Type I error rate arising from multiple item-level comparisons, the Bonferroni adjustment was applied separately to each conceptually related family of comparisons within each use case, rather than across all comparisons in the study. The families comprised 9 subtitle-activation-frequency items and, separately, 13 perceived-usefulness and social-value items in Use Case 1; 9 items in Use Case 2; 4 items in Use Case 3; 5 items in Use Case 4; and 7 items in Use Case 5. Thus, the corresponding Bonferroni-adjusted thresholds were 0.05/9, 0.05/13, 0.05/9, 0.05/4, 0.05/5, and 0.05/7, respectively. Adjusted p-values below 0.05 were considered statistically significant. For mean differences that remained significant after adjustment, Hedges' $g$ was calculated to assess standardized magnitude. Following Cohen's conventional benchmarks for standardized mean differences [30], values of approximately 0.20, 0.50, and 0.80 were interpreted as small, medium, and large effects, respectively.

## C. Sampling Strategy and Achieved Sample

Participants were recruited from an opt-in online panel, professionally managed by Ikerfel S.A., using quota controls for sex and age to approximate the distribution of the target population of Spanish Internet and connected TV users aged 18-64 years. Educational attainment was monitored during fieldwork to avoid marked imbalances in the sample, although it was not used as a formal quota variable. The achieved sample included participants from across Spain.

A total of 1,197 survey sessions were initiated. After excluding incomplete responses and responses that failed the predefined attention or completion-time quality criteria, 626 complete and quality-verified questionnaires were retained for analysis. This corresponds to a completion-and-retention proportion of 52.3% among initiated sessions. Of the excluded sessions, 98% were removed for failing the attention checks, and 2% for inconsistent or implausibly short completion times. The final sample comprised 346 women, 279 men, and one participant who did not disclose their gender. The latter participant was included in whole-sample descriptive statistics but excluded from inferential comparisons between women and men.

[3] https://ikerfel.es/en/ (last access: September 2026)
[4] A copy of the questionnaire draft used for preliminary checking is available at https://tally.so/r/mRDVrP (last access: September 2026).
[5] https://iim.webs.upv.es/Pol%C3%ADtica_de_privacidad.html (in Spanish) (last access: September 2026).
[6] https://www.ibm.com/products/spss-statistics (last access: September 2026)

The resulting sample should be regarded as a non-probability quota sample rather than as a simple random sample. Quota controls, professional panel management, and response-quality screening were used to improve sample balance and data quality, but individual inclusion probabilities were unknown. Accordingly, the inferential analyses are interpreted as model-based comparisons conditional on the achieved sample. They are used to identify and quantify associations and group differences within the study sample and do not account for potential coverage, self-selection, nonresponse, or panel-recruitment error. Thus, the results should not be interpreted as design-based population estimates or as establishing strict statistical representativeness of all Spanish connected TV users.

Each participant evaluated two randomly assigned use cases. The resulting subsample sizes were 242 participants for Use Case 1, 238 for Use Case 2, 245 for Use Case 3, 293 for Use Case 4, and 234 for Use Case 5. Their distribution by gender is reported in Table 1. Because each participant contributed evaluations for two use cases, the five use-case subsamples partially overlapped and were not treated as mutually independent. Inferential analyses were conducted separately within each use case, whereas cross-use-case differences in observed percentages and mean ratings were interpreted descriptively rather than as formal comparisons between independent groups. Further operational details of the survey are provided in Appendix A.

Table 1. Distribution of Participants Assigned to Each Use Case, by Gender

| **Gender** | **UC1** | **UC2** | **UC3** | **UC4** | **UC5** |
|---|---|---|---|---|---|
| **Men (N = 279)** | 100 | 108 | 109 | 141 | 100 |
| **Women (N = 346)** | 141 | 129 | 136 | 152 | 134 |
| **Undisclosed** | 1 | 1 | 0 | 0 | 0 |
| **Total** | **242** | **238** | **245** | **293** | **234** |

### D. *Participant Characteristics*

The sample achieved comprised 626 participants aged 18-64 years ($M$ = 41.67, $SD$ = 11.71). Women accounted for 55.27% of the sample ($n$ = 346), men for 44.57% ($n$ = 279), and one participant did not disclose their gender. Regarding educational attainment, 3% reported primary education or lower, 47% reported secondary education or vocational training, 39% reported a university degree, and 12% reported postgraduate education. Most participants were employed (72.5%), followed by unemployed participants with previous work experience (10.4%), university students (6.9%), homemakers (5.1%), and retired participants or those receiving permanent disability benefits (3.7%). The age distribution is reported in Table 2.

Table 2. Age Distribution

| **Age Cohort (Yrs)** | **Total Cluster** |
|---|---|
| 18–24 | 63 (10.06%) |
| 25–34 | 130 (20.77%) |
| 35–44 | 163 (26.04%) |
| 45–54 | 159 (25.40%) |
| 55–64 | 111 (17.73%) |
| Total | 626 (100.00%) |

## VI. Survey Results and Statistical Analysis

This section presents the structured empirical findings derived from the quantitative analysis of the collected survey data. The presentation is organized into a macro-analysis of the general technological profiles, followed by localized inferential metrics for each evaluated use case. For all item-level comparisons of mean scores by gender, the participant with undisclosed gender was excluded, and two-tailed Welch's independent-samples t-tests were used. Bonferroni adjustment was applied separately within each conceptually related family of comparisons in each use case, as specified in Section V-B; in Use Case 1, the nine behavioral items and the 13 perceived-usefulness and social-value items were treated as two distinct families.

### A. *General Results*

A1) *Technology Profile and XR Familiarity.*

Participants were asked to self-identify their technological profile. Only 7.7% (16 women, 32 men) reported having technical training or professional experience in engineering or related fields (computers, electronics, etc.), while 32.3% (105 women, 97 men) identified themselves as users of similar technologies without formal expertise. The majority (60.1%; 225 women, 150 men) did not consider themselves particularly related to immersive technologies. Men have a slightly higher percentage of technical profiles and technology users, a difference that is statistically significant (p=0.001 in the $\chi^2$ test).

When assessing familiarity with AR/VR technologies (5-point Likert scale: 1-Very low; 2-Low; 3-Medium; 4-High; and 5-Very high), Table 3 shows that 16.1% reported high or very high familiarity, 46% reported medium familiarity, and 37.9% reported low or very low familiarity. Familiarity levels differed significantly by gender, $\chi^2(4, N = 625) = 27.38$, $p <$ 0.001. Two-tailed pairwise comparisons of column proportions using $z$-tests with Bonferroni-adjusted $p$-values showed that women were more likely than men to report medium familiarity (51.7% vs. 38.7%, adjusted $p = 0.0058$). In contrast, men were more likely to report high familiarity than women (19.7% vs. 7.8%, adjusted $p < 0.001$). Although men also reported very high familiarity more frequently than women (4.7% vs 1.7%), this difference was no longer statistically significant after the Bonferroni adjustment (adjusted $p = 0.171$).

Table 3. Self-reported familiarity with AR/VR technologies, by gender*

| Familiarity level | Women (n=346), n (%) | Men (n=279), n (%) | TOTAL (*N*=626), *n* (%) |
|---|---|---|---|
| Very low | 38 (11.0%)[a] | 25 (9.0%)[a] | 63 (10.1%) |
| Low | 96 (27.7%)[a] | 78 (28.0%)[a] | 174 (27.8%) |
| Medium | 179 (51.7%)[a] | 108 (38.7%)[b] | 288 (46.0%) |
| High | 27 (7.8%)[a] | 55 (19.7%)[b] | 82 (13.1%) |
| Very high | 6 (1.7%)[a] | 13 (4.7%)[a] | 19 (3.0%) |
| Total | 346 (100%) | 279 (100%) | 626 (100%) |

* Different superscript letters (*a* and *b*) within a row indicate a statistically significant difference between women and men, based on two-tailed pairwise comparisons of column proportions using z-tests with Bonferroni-adjusted p-values

Although 42.3% (138 women, 127 men) of participants reported prior AR/VR experience, these experiences were predominantly entertainment-driven (e.g., AR/VR-based videogames or AR/VR-based apps in museums or theme parks), and responses often revealed conceptual confusion between AR and VR. Thus, participants largely evaluated immersive TV services based on expectations rather than on consolidated prior use. No statistically significant differences were observed between genders (p=0.18>0.05 in the $\chi^2$ test).

*A2) Second-Screen Usage Habits.*

Second-screen usage during TV consumption is widespread: 83.1% of participants (289 women, 230 men) reported using an additional device while watching TV. 4.8% always use them; 27.2% use them quite frequently; 32.1% use them in some cases, depending on the content and genre; and 19% use them very rarely. Both genders show the same distribution of secondary device use frequency (p=0.995 > 0.05 in the $\chi^2$ test). The smartphone is the dominant companion device (91.5%), followed by laptops (37.7%), tablets (33.7%), and desktop PCs (10%). In contrast, AR (1%) and VR (2.3%) HMDs are rarely used in current TV consumption contexts. These findings confirm that multi-device ecosystems are already normalized in domestic environments. However, integrating AR/VR devices would represent a qualitative shift in interaction paradigms rather than a simple extension of current second-screen practices.

*A3) Preferences for Multisensory Augmentation.*

A substantial majority of participants expressed interest in experiencing multisensory effects while watching TV content (79.4%; 273 women and 223 men). No statistically significant association was observed between gender and general interest in multisensory augmentation, $\chi^2(1, N = 625) = 0.10$, $p = 0.753$. Among the 496 interested participants, the most frequently selected effects were ambient lighting (78.2%; 216 women and 172 men), seat vibration or movement (63.3%; 168 women and 146 men), wind effects (54.2%; 132 women and 137 men), pleasant scents (53.6%; 164 women and 102 men), heating effects (49.2%; 127 women and 117 men), cooling effects (47.2%; 114 women and 120 men), fog effects (40.1%; 107 women and 92 men), any type of scent, including both pleasant and unpleasant scents (40.1%; 96 women and 103 men), and water spray (32.7%; 90 women and 72 men). Statistically significant gender associations were found for pleasant scents, $\chi^2(1, N = 496) = 10.14$, $p = 0.001$, and wind effects, $\chi^2(1, N = 496) = 8.46$, $p = 0.004$. Pairwise comparisons of column proportions with Bonferroni adjustment across the nine sensory effects confirmed that women selected pleasant scents more frequently than men (60.1% vs. 45.7%, adjusted $p = 0.013$). In contrast, men selected wind effects more frequently than women (61.4% vs. 48.4%, adjusted $p = 0.033$). No other gender differences remained statistically significant after adjustment.

Regarding the content genres for which participants expressed interest in multisensory effects, the most frequently selected categories were films and series (95.4%; 262 women and 211 men), documentaries (53.8%; 142 women and 125 men), music videos (39.3%; 115 women and 80 men), sports events (35.1%; 60 women and 114 men), games and quiz shows (28.0%; 72 women and 67 men), educational programs (23.4%; 56 women and 60 men), reality shows (16.3%; 57 women and 24 men), news (13.1%; 32 women and 33 men), award galas and similar programs (9.1%; 25 women and 20 men), and cooking programs (1.4%; 5 women and 2 men). Statistically significant gender associations were found for sports events, $\chi^2(1, N = 496) = 45.77$, $p < 0.001$, and reality shows, $\chi^2(1, N = 496) = 9.19$, $p = 0.002$. Bonferroni-adjusted pairwise comparisons of column proportions confirmed that men selected sports events more frequently than women, whereas women selected reality shows more frequently than men. No other gender differences remained statistically significant after adjustment. Overall, these findings indicate a particularly strong interest in multisensory augmentation for narrative-driven and cinematic audiovisual content.

## B. Localized Results per Specific Use Case

In this section, the results obtained for each use case are summarized.

*B1) Use Case 1 – Adaptive and Personalized Subtitles.*

This use case enables users to customize subtitle parameters, including language, font size, and spatial positioning, on either the primary TV display or a secondary companion device. The sample size for this use case was 242 participants (Table 1).

*Participants' previous Experience.* This use case recorded the highest rates of both prior user experience and participant interest across all evaluated scenarios (Table 4). Gender segmentation revealed no statistically significant differences between male and female participants regarding either prior experience with this scenario ($\chi^2$ test, $p = 0.18 > 0.05$) or explicit interest in adopting the service ($p = 0.85 > 0.05$).

Specifically, 38.8% of respondents ($n = 94$) reported prior experience with subtitle customization, predominantly on primary displays via Over-The-Top (OTT) streaming platforms such as Netflix and Prime Video. In contrast, only 17.0% had interacted with subtitles on secondary companion devices. Furthermore, the general assessment among experienced participants ($n = 94$) was predominantly favorable: 75.5% evaluated their experience positively ($n = 71$; comprising 28 *Very Positive* and 43 *Positive* ratings), 23.4% remained neutral ($n = 22$), and only 1.1% ($n = 1$) reported a *Negative* experience, with no *Very Negative* responses recorded. Qualitative feedback from the single negative respondent, along with two participants who reported positive overall experiences, highlighted minor technical issues, predominantly related to synchronization.

The most used secondary screen for viewing subtitles is the smartphone (21 participants), followed by the tablet (9), the laptop (7), and VR (2) and AR (1) HMDs.

*Participants' Behavior:* Following the statistical analysis framework outlined in Section V-B, subtitle activation frequency was evaluated across TV content genres on a 4-point Likert scale (1 = *Never* to 4 = *Always*). Activation levels varied substantially by genre, peaking in films and series ($M$ = 2.38) and remaining lowest in sports ($M$ = 1.42) and news ($M$ = 1.52).

Table 4. Similar previous experiences and interest in use cases

| Use case | Percentage of participants (%) | |
|---|---|---|
| | With similar experiences | Interested in this use case |
| 1. Adaptive and Personalized Subtitles | 38.9% | 76.86% |
| 2. Complementary (broadband) content displayed on AR HMD, providing additional information synchronized with the main (broadcast) TV program. | 2.1% | 64.71% |
| 3. Virtual Large Screen via AR HMDs. | 0% | 66.94% |
| 4. Combination of traditional (broadcast) 2D TV content with synchronized virtual 3D objects/models/animations presented in secondary devices, enabling spatial visualization of complementary information. | 14% | 69.97% |
| 5. Combination of traditional (broadcast) TV content with complementary (broadband) 360° video | 15.38% | 66.67% |

Two-tailed Welch's independent-samples t-tests indicated nominal differences for films and series (women $M$ = 2.26, men $M$ = 2.54, $p$ = 0.022), documentaries (women $M$ = 1.91, men $M$ = 2.19, $p$ = 0.026), and sports (women $M$ = 1.32, men $M$ = 1.57, $p$ = 0.021). However, none of the nine comparisons remained statistically significant after applying the Bonferroni adjustment to this behavioral family (adjusted p = 0.201, 0.237, and 0.190, respectively; all other adjusted p-values = 1.000). Therefore, the data do not provide evidence of gender differences in subtitle activation frequency for any of the nine content genres after controlling for the familywise error rate.

*Perceived Usefulness and Social Value.* Participants provided their level of agreement with the following statements regarding this use case by using a 5-point Likert scale (1-*Totally disagree*, 2-*Partially disagree*, 3-*Neutral*, 4-*Partially agree*, 5-*Totally agree*):

*Q1. In general, I find the service/functionality described in this use case useful.*

*Q2. Customizing the subtitle format (size, font, color, position) is a useful and interesting feature for me.*

*Q3. Customizing the subtitle format (size, font, color, position, etc.) may contribute to the social inclusion of people with visual impairments.*

*Q4. Selecting a subtitle language may contribute to the social inclusion of people who do not understand the AV content's original language (e.g., foreigners).*

*Q5. The integration, synchronization, and subtitle format customization on secondary devices (tablets, smartphones, laptops, or AR HMDs) are useful and interesting features for me.*

*Q6. The ability to configure a time lag (positive or negative) for subtitle display relative to the AV content (e.g., to allow more time to read them before they disappear) is a useful feature.*

*Q7. This type of service/platform is useful in environments where people speaking different languages are present (e.g., apartments with international students, visits from foreign relatives, etc.).*

*Q8. This type of service/platform is useful in environments where people with some form of AV impairment (e.g., hearing, visual, etc.) are present.*

*Q9. This type of service/platform is useful in crowded environments where audio cannot be heard, or a high volume is not advisable (e.g., restaurants, museums or hospitals).*

*Q10. This type of service/platform is useful for language learning.*

*Q11. The simultaneous selection of multiple languages for subtitles is a useful and interesting feature.*

*Q12. Displaying multiple subtitle lines simultaneously is a useful and interesting feature (for example, giving more time to read the subtitles before they disappear).*

*Q13. Suitable technology for displaying subtitles on secondary devices (e.g., tablets, smartphones, laptops, AR HMDs), such as that in the use case, solves a real problem.*

Figure 8 shows the distribution of responses for each statement, and Table 5 summarizes their statistical values. They received consistently high scores (mean values ranged from 3.86 to 4.58). Highest-rated aspects included: contribution to inclusion of non-native speakers (Q4, M=4.58); contribution to inclusion of people with AV impairments (Q3, M=4.52), 77.3% consider it as very useful for integrating individuals with hearing impairments or non-native speakers; usefulness for language learning (Q10, M=4.46), identified as a major benefit by 74% of respondents; usefulness in noisy environments (Q9, M=4.42), it enables silent TV consumption without disturbing others (59.5%). Participants also agreed that this service/technology 'solves a real problem' (Q13, M=4.04). Overall, this use case was perceived as socially valuable, functionally useful, and technologically plausible.

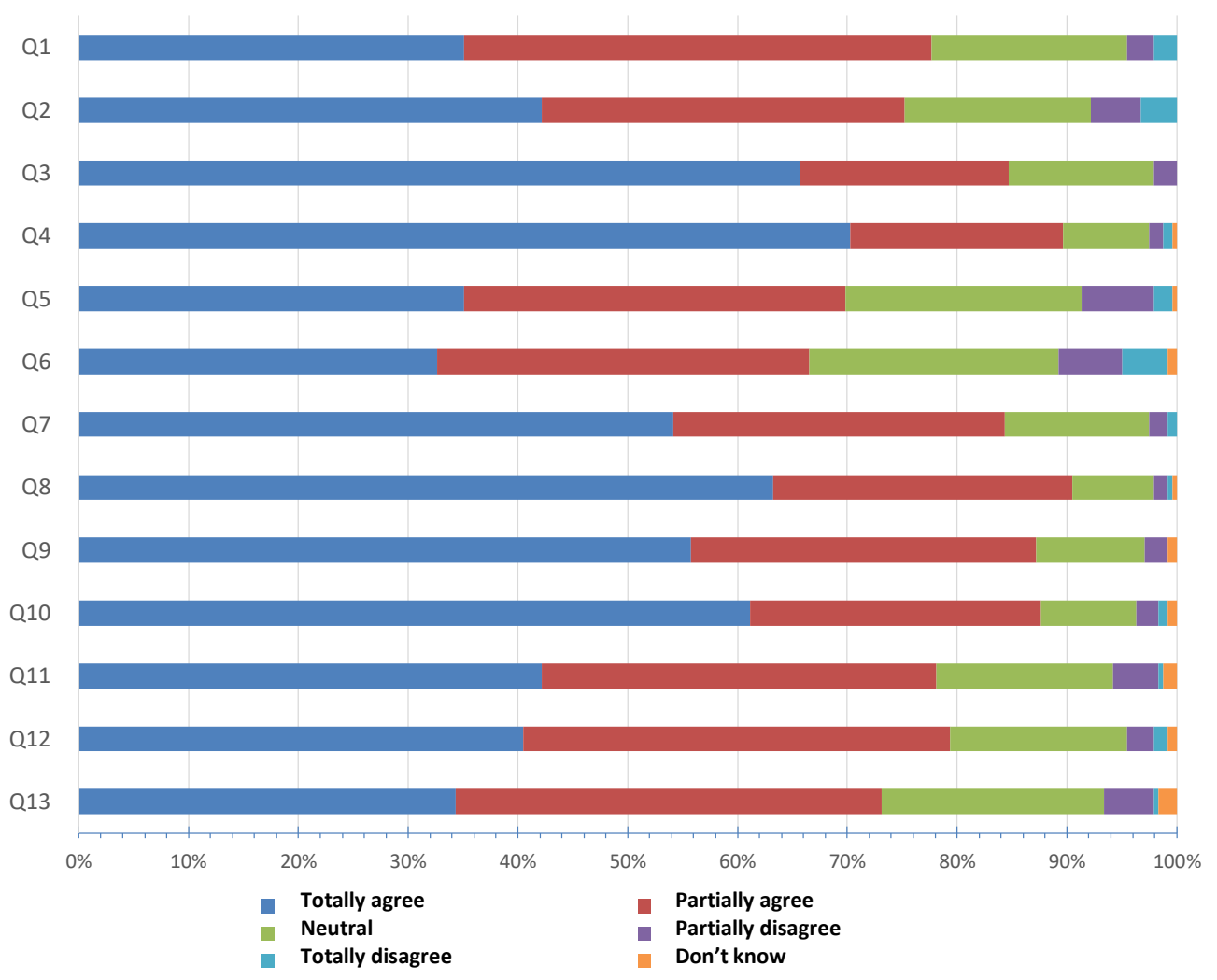

Figure 8. Participants' answers about statements regarding use case 1

Table 5. Statistical values of the responses about statements regarding use case 1

| Statement | Mean value | Standard Deviation |
|---|---|---|
| **Q1** | 4.06 | ±0.902 |
| **Q2** | 4.06 | ±1.035 |
| **Q3** | 4.48 | ±0.801 |
| **Q4** | 4.58 | ±0.761 |
| **Q5** | 3.95 | ±0.993 |
| **Q6** | 3.86 | ±1.073 |
| **Q7** | 4.35 | ±0.833 |
| **Q8** | 4.52 | ±0.725 |
| **Q9** | 4.42 | ±0.756 |
| **Q10** | 4.46 | ±0.807 |
| **Q11** | 4.17 | ±0.878 |
| **Q12** | 4.16 | ±0.873 |
| **Q13** | 4.04 | ±0.883 |

Responses were analyzed according to participants' gender using two-tailed Welch's independent-samples t-tests, excluding the participant with undisclosed gender. Bonferroni adjustment was applied across the 13 conceptually related perceived-usefulness and social-value items, separately from the nine behavioral comparisons reported above. Q13 was the only item that remained statistically significant after adjustment: men agreed more strongly than women that the technology presented in Use Case 1 solves a real problem (women M = 3.90, men M = 4.24), Welch's t (233.99) = -3.14, raw p = 0.0019, Bonferroni-adjusted p = 0.0245. The standardized difference was small-to-medium according to Cohen's conventional benchmarks (Hedges' g = 0.40, calculated as men minus women). No other item in this 13-comparison family was statistically significant after adjustment (all adjusted p-values = 1.000).

*Benefits of the use case*. Regarding the benefits of this use case (Table 14), only 2 participants (0.8% of the sample) reported no benefits to the described use case. The rest contributed one or more benefits. The main benefits indicated were the contribution to social integration (77%); allowing for a more engaging and personalized experience (76%); learning other languages (74%); and watching TV without sound in the room (59.5%).

*Limitations of the use case.* Regarding the weaknesses of this use case (Table 15), 33 participants (13.6%) stated they did not see any limitations. The rest reported at least one limitation, highlighting technical issues, such as those related to the lack of synchronization of the subtitles with the AV content (57.4%) or when the Internet connection does not work or is not fast enough and the subtitles are not downloaded on time (55%); as well as distraction or lack of concentration on the AV content (34.7%) or the discomfort or annoyance caused by reading them in a virtual space outside the TV frame when using AR HMDs (59.5%). In cases where secondary devices are used for viewing subtitles, the cost of these devices (31.8%) and the inconvenience of reading subtitles on them (29.3%) are also considered limitations.

*Participants' interest in this use case.* Participants were asked whether, if this service/use case were available on their home TV, they had the necessary knowledge and the appropriate devices, they would be interested in an experience like the one described in the use case. On the one hand, 186 participants (76.86% of the sample in this use case) responded affirmatively (Table 4). They mainly would use it for viewing different types of content (movies, series, documentaries, programs, news, reality shows...) with customized subtitles (31.2%); for studying languages (19.4%); and for enjoying content in its original version and in other languages (16.1%). The preferred secondary devices are smartphones (62.4%) and AR HMDs (54.8%). On the other hand, 27 participants (11.16%) responded negatively, and the rest were undecided.

Participants did not provide any additional open-ended suggestions for improving this use case beyond the functionality preferences collected in the questionnaire. Overall, it was perceived as socially valuable, functionally useful, and technologically plausible.

*B2) Use Case 2 – Complementary (broadband) content displayed on AR HMD, providing additional information synchronized with the main (broadcast) TV program.*

This use case allows users to access synchronized additional content (e.g., statistics, sign language interpretation videos, or alternative camera angles) through AR devices. The sample size for this use case was 238 participants (Table 1).

*Participants' previous Experience.* Only a very small percentage (5 participants, 2.1%) reported having any previous experience like this case (Table 4), but none with AR HMDs, only with tablets or smartphones. No technical problems were reported. Two participants indicated they had a *Positive* experience; two responded *Neutral*; and only one participant reported a *Negative* experience, with a negative feeling related to an *'intrusive'* experience without providing further details.

*Perceived Usefulness and Social Value.* Participants provided their level of agreement with the following statements regarding this use case:

*Q1. In general, I find the service/functionality described in this use case useful.*

*Q2. The integration, synchronization, and customization of additional video content playout, complementary to the content on the main TV are useful and interesting features.*

*Q3. This type of service/platform is useful in environments where people speak different languages (e.g., apartments with foreign students, visits from foreign relatives, etc.).*

*Q4. This type of service/platform is useful for environments where people with some type of audiovisual impairment (e.g., hearing or visual impairment) are present.*

*Q5. The service/functionality described in the use case helps increase user attention while consuming TV content.*

*Q6. I believe the use case described provides a greater sense of engagement and makes the content on the main TV more attractive (i.e., it keeps the user watching the TV content without changing the channel).*

*Q7. A suitable technology for displaying additional video content that complements the main TV content, to be played on secondary devices (e.g., tablets, smartphones, laptops, AR HMDs) may solve a real problem.*

*Q8. When the playout of complementary content on secondary devices is out of sync (ahead or behind) with respect to the main TV content (i.e., a lack of synchronization), the TV consumption experience is unsatisfactory.*

*Q9. Connected TV consumption is mostly done together with other people. Suppose you are watching augmented virtual TV at the same time another family member is watching the same content on the main TV. The lack of synchronization between the playout of supplementary content on secondary devices and the playout of the main TV content prevents both of you from having a satisfactory shared TV consumption experience (e.g., one in which you both comment on what you are watching at any given moment).*

Figure 9 shows the distribution of responses for each statement, and Table 6 summarizes their statistical values. The sample of Use Case 2 clearly aligns with all the statements. All of them received consistently high scores (mean values ranged from 3.8 to 4.43). Highest-rated aspects included contribution to the inclusion of people with audiovisual impairments (Q4, M = 4.43), contribution to the inclusion of people who speak other languages (Q3, M = 4.33), and the importance of synchronization for satisfactory individual and shared viewing experiences (Q8-Q9, M = 4.13 and 4.08, respectively). The integration, synchronization, and customization of complementary video playout were also rated positively (Q2, M = 4.11), and participants agreed that the service may solve a real problem (Q7, M = 3.90). Gender comparisons used two-tailed Welch's independent-samples t-tests. After Bonferroni adjustment across the nine items in this use-case family, none of the differences between women and men was statistically significant (all Bonferroni-adjusted p-values = 1.000 after capping at 1.000; raw p-values ranged from 0.173 to 0.822). Overall, this use case was perceived as socially valuable, functionally useful, and technologically plausible, with no evidence of item-level gender differences after adjustment for multiplicity.

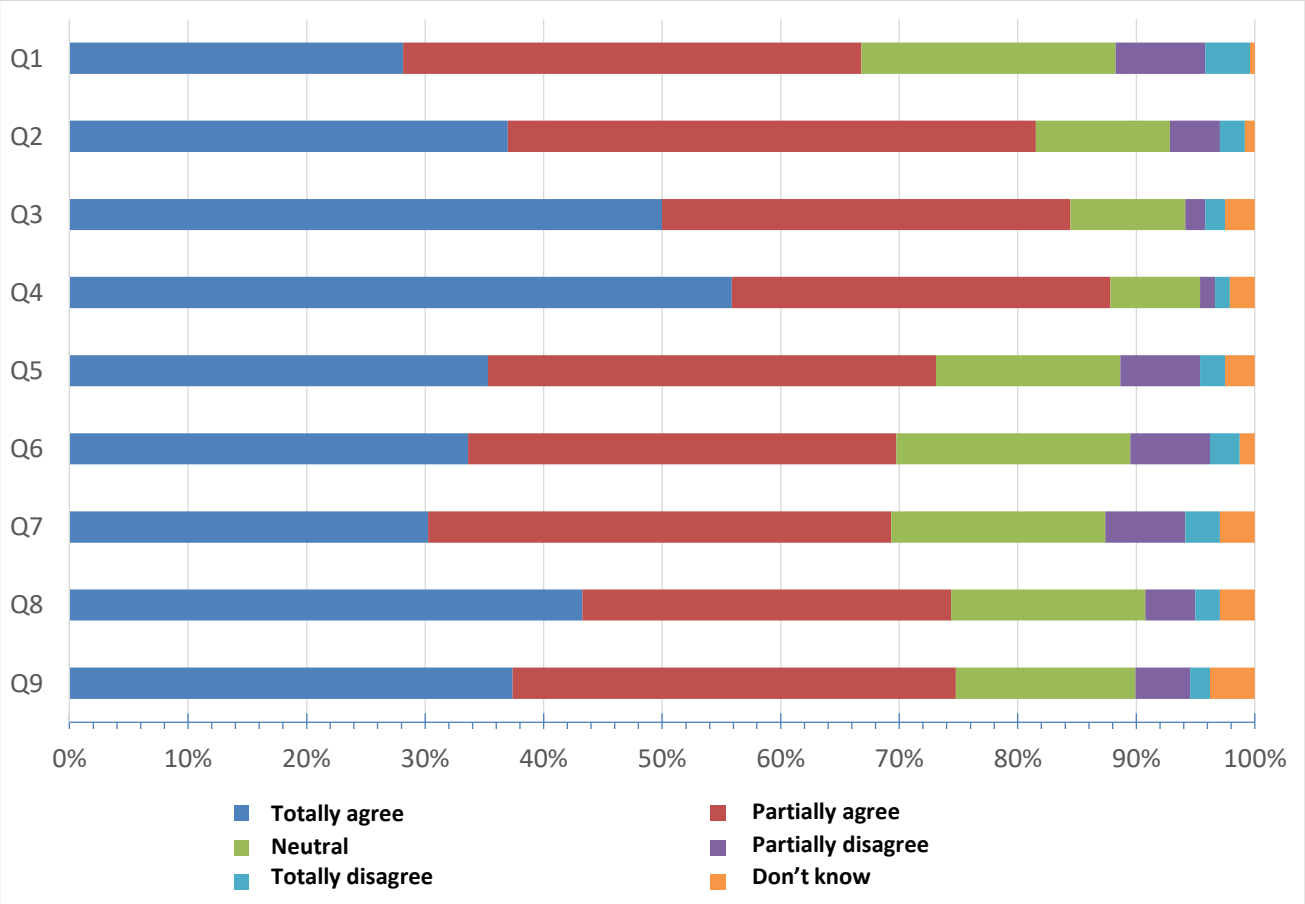

Figure 9. Participants' answers about statements regarding use case 2

Table 6. Statistical values of the responses about statements regarding use case 2

| Statement | Mean value | Standard Deviation |
|---|---|---|
| **Q1** | 3.80 | ±1.053 |
| **Q2** | 4.11 | ±0.916 |
| **Q3** | 4.33 | ±0.856 |
| **Q4** | 4.43 | ±0.796 |
| **Q5** | 4.00 | ±0.998 |
| **Q6** | 3.93 | ±1.021 |
| **Q7** | 3.90 | ±1.020 |
| **Q8** | 4.13 | ±0.985 |
| **Q9** | 4.08 | ±0.945 |

*Benefits of the use case*. Regarding the benefits of this use case (Table 14), only 11 participants (4.6% of the sample of this use case) reported no benefits to the use case. The rest have provided at least one benefit. The greatest benefits considered are social integration (70.6%), and the possibility of enjoying a more attractive and personalized experience (65.5%), learning other languages (52.1%), being able to access complementary video content can hook (i.e., engage) them to continue watching the content of the tuned TV channel (46.6%), as well as obtaining greater knowledge/information about the main broadcast content watched on the TV main device (44.1%).

*Limitations of the use case.* Regarding the weaknesses of this use case (Table 15), 17 participants (7.1%) reported seeing no limitations. The rest have declared at least one limitation, highlighting technical inconveniences, such as when the Internet connection does not work or is not fast enough and the complementary content is not downloaded on time (60.9%) or those related to the lack of synchronization between the playout of the complementary content and that of the main content on the TV (59.7%); as well as the cost of secondary devices (48.7%), and the discomfort (36.1%) or distraction (32.4%) caused by viewing complementary content on secondary devices. It should be noted that 53 participants (22.3%) have pointed out the problem that using battery-powered devices can pose and the problems that this entails, 51 participants (21.4%) have indicated the lack of technological training as a limitation in this use case, 45 participants (18.9%) see possible visual harmful effects, and 39 participants (16.4%) believe that the technology is not yet mature enough.

*Participants' interest in this use case.* Participants were asked whether, if this service/use case were available on their home TV, they had the necessary knowledge and the appropriate devices, they would be interested in an experience like the one described in the use case. 154 participants (64.71% of the sample) responded affirmatively (Table 4). They mainly would use it to view different types of content, such as movies, series, documentaries, programs, news, or reality shows (37.7%); to watch sports content (16.9%); and to enjoy content in its original version and in other languages (10.4%). Those participants were also asked to indicate the TV programs or genres for which they believed supplementary video content would be most useful. The responses are presented in Table 7. They would find it most useful while watching movies (72.7%), series (57.8%), or sports (36.4%). As shown in the table, men cited sports and educational programs more frequently, while women cited reality shows more often. These differences were statistically significant according to the $\chi^2$ test ($p < 0.001$), with post-hoc pairwise comparisons of proportions (z-tests with Bonferroni correction) confirming the observed gender differences.

If their costs were not a problem for participants (they were asked not to consider them), their preferred secondary devices for this use case were AR HMDs (71.4%), smartphones (61.7%), laptops (48.1%) and tablets (40.9%).

Table 7. TV genres for which participants believed supplementary video content would be most useful (multiple choice question)

| Content genre | Responses by gender** | | Responses* |
|---|---|---|---|
| | Women | Men | Total |
| **Films** | 60 (69.8%) | 52 (76.5%) | 112 (72.7%) |
| **Series** | 54 (62.8%) | 35 (51.5%) | 89 (57.8%) |
| **Sports** | 19 (22.1%) | 37 (54.4%) | 56 (36.4%) |
| **Musical Videos** | 30 (34.9%) | 21 (30.9%) | 51 (33.1%) |
| **TV reports** | 24 (27.9%) | 19 (27.9%) | 43 (27.9%) |
| **Education** | 16 (18.6%) | 24 (35.3%) | 40 (26.0%) |
| **Cultural programs** | 17 (19.8%) | 20 (29.4%) | 37 (24.0%) |
| **Music (with lyrics)** | 18 (20.9%) | 17 (25.0%) | 35 (22.7%) |
| **News** | 14 (16.3%) | 17 (25.0%) | 31 (20.1%) |
| **All types of content** | 16 (18.6%) | 13 (19.1%) | 29 (18.8%) |
| **Games/quizzes** | 14 (16.3%) | 14 (20.6%) | 28 (18.2%) |
| **Events (galas…)** | 8 (9.3%) | 10 (14.7%) | 18 (11.7%) |
| **Reality shows** | 14 (16.3%) | 3 (4.4%) | 17 (11.0%) |
| **No fiction** | 4 (4.7%) | 1 (1.5%) | 5 (3.2%) |
| **Documentaries** | 2 (2.3%) | 0 (0.0%) | 2 (1.3%) |
| **Advertisements** | 1 (1.2%) | 0 (0.0%) | 1 (0.6%) |
| **Total.-** | 86 | 68 | 154 (100%) |

* % regarding 154 participants
** regarding the number of women and men (86 and 68, respectively).

Although more than 64% of participants showed active interest in this use case as an augmented TV service, it was valued primarily as optional and complementary rather than as a replacement for traditional TV viewing.

Regarding participants' suggestions, some of them pointed out the need for accessible, attractive and friendly interfaces with easy and natural interaction (e.g., also with gestures and voice control), the need for synchronization and for cheaper and more comfortable AR HMDs.

*B3) Use Case 3 – Virtual Large Screen via AR HMDs.*

This use case allows users to view TV content through AR HMDs as if displayed on a large-scale virtual TV screen in front of them. Users with visual impairments can view larger TV images that better suit their needs. The sample size for this use case was 245 participants (Table 1).

*Participants' previous Experience.* No participants reported any prior experience related to this case (Table 4).

*Perceived Usefulness and Social Value.* The participants also provided their level of agreement with the following statements regarding this use case:

*Q1. The service/functionality described in the use case is useful and interesting; it provides added value to the TV content consumption experience.*
*Q2. The service/functionality described is especially useful in environments where users have visual impairments.*
*Q3. Connected TV consumption is mostly done in company. Suppose you are watching augmented virtual TV on your AR HMDs at the same time another family member is watching the same content on the main TV. The lack of synchronization between the playout of the augmented virtual TV content and the main TV content prevents both of you from having a satisfactory shared TV viewing experience (for example, one in which you both comment on what you are watching at any given moment).*
*Q4. I believe the described service/functionality solves a real problem.*

Figure 10 shows the distribution of responses for each statement, and Table 8 summarizes their statistical values. The sample for Use Case 3 showed high agreement with the statements (mean values ranging from 3.90 to 4.47). Most participants considered the service useful and interesting, and believed it added value to TV content consumption, especially for users with visual impairments. The lack of synchronization between the virtual enlarged TV content and the main TV content was also perceived as potentially detrimental to a satisfactory shared viewing experience. Gender comparisons used two-tailed Welch's independent-samples t-tests. After Bonferroni adjustment across the four items in this use-case family, none of the differences between women and men was statistically significant (adjusted p-values = 0.680 for Q3 and 1.000 for Q1, Q2, and Q4; raw p-values ranged from 0.170 to 0.953).

Overall, the concept of projecting a large virtual screen via AR HMDs generated curiosity and interest, particularly among younger, more technologically-savvy users.

*Benefits of the use case.* Regarding the benefits of this use case (Table 14), only 16 participants (6.5% of the sample) reported no benefits. The rest have selected or contributed at least one benefit. 70.2% highlighted the enhanced visual experience due to a better visualization of contents on a large-scale display; help for users with visual impairments, which contributes to their social integration (59.2%); a more attractive and personalized experience (56.3%); and improved viewing conditions in constrained spaces, it allows large-screen viewing regardless of the physical size of the actual TV set or the room (53.9%).

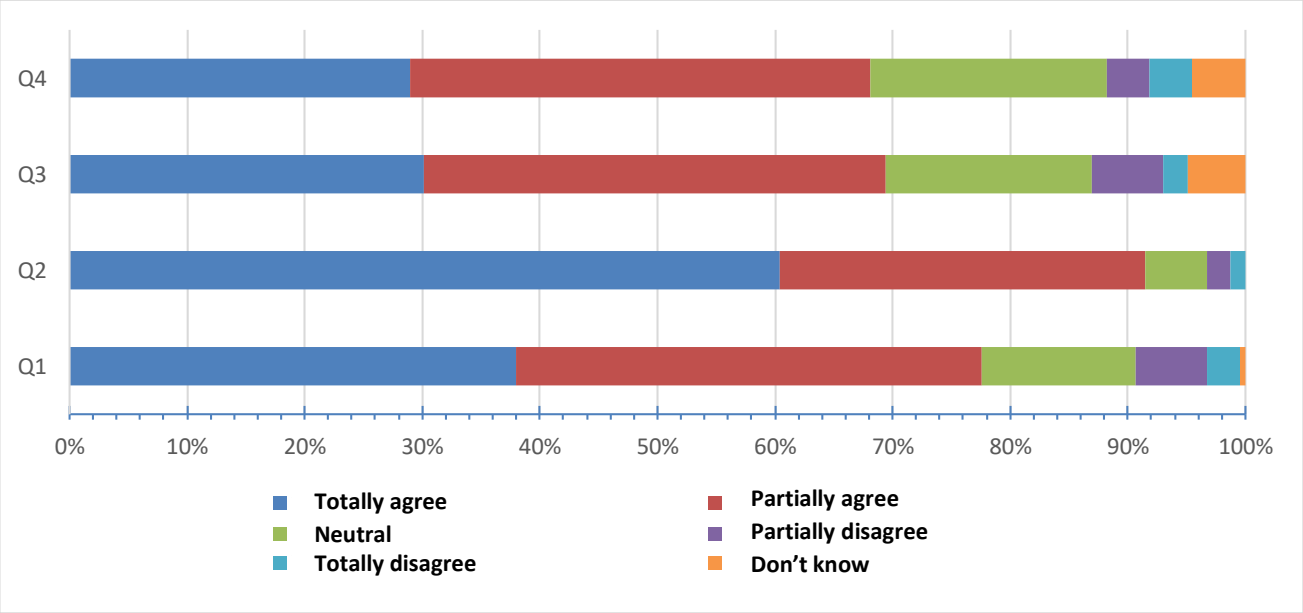

Figure 10. Participants' answers about statements regarding use case 3

Table 8. Statistical values of the responses about statements regarding use case 3

| Statement | Mean value | Standard Deviation |
|---|---|---|
| **Q1** | 4.04 | ±1.009 |
| **Q2** | 4.47 | ±0.792 |
| **Q3** | 3.94 | ±0.976 |
| **Q4** | 3.90 | ±1.004 |

*Limitations of the use case.* Regarding the weaknesses of this use case (Table 15), 25 participants (10.2%) reported seeing no limitations. The rest have declared at least one limitation, highlighting cost (57.6%); in shared consumption sessions, when interacting with more people watching TV in the same room, the lack of synchronization of the virtual (AR-based) TV content with the main TV content (49.4%); comfort and ergonomic concerns (40%); lack of the required technical skills or training to use the AR HMDs and the service (27.3%); problems that battery-powered devices can pose and other problems they entail (24.9%); possible visual fatigue or harmful effects (22.9%); and that the technology is not yet mature enough (18%).

*Participants' interest in this use case.* Participants were asked whether, if this service/use case were available on their home TV, they had the necessary knowledge and the appropriate devices, they would be interested in an experience like the one described in the use case. 164 participants (66.94% of the sample) responded affirmatively (Table 4). They mainly would use it for watching all types of TV content (54.9%); for watching sports events (12.8%); for having a more immersive TV consumption experience (11.6%); or for virtually enlarging their TV screen (3%); and for enjoying content in its original version and in other languages (10.4%). 13.4% reported that people with visual impairments would use it always, replacing the TV device. Those participants were also asked to indicate which TV programs or genres this use case would be most useful for. The responses are presented in Table 9. They would find it most useful while watching movies (66.5%), followed by series (48.8%) and sports (32.3%). As shown in the table, men cite sports programs more frequently, which is the only statistically significant difference between the two genders (z-tests with Bonferroni correction).

Table 9. TV genres for which participants believed the use of large virtual screens through AR HMDs would be most useful (multiple choice question)

| Content genre | Responses by gender** | | Responses* |
|---|---|---|---|
| | Women | Men | Total |
| **Films** | 59 (67.0%) | 50 (65.8%) | 109 (66.5%) |
| **Series** | 42 (47.7%) | 38 (50.0%) | 80 (48.8%) |
| **Sports** | 22 (25.0%) | 31 (40.8%) | 53 (32.3%) |
| **Musical Videos** | 29 (33.0%) | 17 (22.4%) | 46 (28.0%) |
| **All types of content** | 22 (25.0%) | 22 (28.9%) | 44 (26.8%) |
| **TV reports** | 25 (28.4%) | 18 (23.7%) | 43 (26.2%) |
| **Cultural programs** | 15 (17.0%) | 16 (21.1%) | 31 (18.9%) |
| **Education** | 14 (15.9%) | 12 (15.8%) | 26 (15.9%) |
| **Games/quizzes** | 12 (13.6%) | 12 (15.8%) | 24 (14.6%) |
| **News** | 7 (8.0%) | 9 (11.8%) | 16 (9.8%) |
| **Events (galas…)** | 10 (11.4%) | 4 (5.3%) | 14 (8.5%) |
| **Music (with lyrics)** | 8 (9.1%) | 6 (7.9%) | 14 (8.5%) |
| **Reality shows** | 8 (9.1%) | 4 (5.3%) | 12 (7.3%) |
| **No fiction** | 3 (3.4%) | 2 (2.6%) | 5 (3.0%) |
| **Documentaries** | 0 (0.0%) | 2 (2.6%) | 2 (1.2%) |
| **Cooking programs** | 0 (0.0%) | 1 (1.3%) | 1 (0.6%) |
| **Total.-** | 88 | 76 | 164 (100%) |

* % regarding 164 participants
** regarding the number of women and men (88 and 76, respectively).

When asked for suggestions, 11 participants (4.5%) were concerned about the possible visual harmful effects; the comfort of wearing the AR HMDs for long periods of time (8, 3.3%); accessibility issues (5, 2%); and image quality, sharpness, colors, etc. (2, 0.8%).

In summary, acceptance of this service/use case appears conditional on device maturity and usability improvements. Anyway, more than 65% of users expressed interest in this kind of experience.

*B4) Use Case 4 – Combination of traditional (broadcast) 2D TV content with synchronized virtual 3D objects/models/animations presented in secondary devices, enabling spatial visualization of complementary information.*

This use case allows users to interact with TV content-related 3D objects (e.g., products or characters) or VR animations of products advertised on TV (e.g., zoom in/out, rotate, move, etc.) rendered on AR/VR devices. Users with visual impairments can view certain 3D objects and products at a larger size that is more appropriate for them. The sample size for this use case was 293 participants (Table 1).

*Participants' previous Experience.* 41 participants (14%) reported prior experience with VR, AR, or 3D, but very few reported this experience in the context of the use case (Table 4). Only 12 participants used AR HMDs to view virtual 3D objects in front of them. Two of them reported using AR apps in museums to view 3D replicas of ancient objects, and one reported using IKEA AR apps[7] in decoration/design shops to see how the furniture would look. The other participants did not provide any details. All but one reported positive experiences. That participant reported a negative experience because the system was not sufficiently realistic and the image quality was inadequate.

*Perceived Usefulness and Social Value.* Participants provided their level of agreement with the following statements regarding this use case:

*Q1. In general, I find the service/functionality described in this use case useful.*
*Q2. The service/functionality described in the use case provides added value to TV content consumption.*
*Q3. This type of service/functionality is useful for marketing campaigns where advertisers can make 3D models of products available online so users can view them on 3D viewers at home.*
*Q4. The service/functionality described in the use case provides the user with a greater sense of engagement and makes the TV content playing on the main TV more appealing (i.e., it keeps the user watching the TV content without changing the channel).*
*Q5. Suitable technology for displaying additional 3D content that complements the main TV content, to be played on secondary devices (e.g., tablets, smartphones, laptops, AR HMDs), solves a real problem.*

Figure 11 shows the distribution of responses for each statement, and Table 10 summarizes their statistical values. The sample for Use Case 4 also aligned with the statements, with mean values ranging from 3.85 to 4.33. Participants considered the service useful (Q1, M = 3.88), particularly for marketing campaigns in which advertisers make interactive 3D product models available for home viewing (Q3, M = 4.33). They also agreed that the service may increase engagement (Q4, M = 4.02) and may solve a real problem (Q5, M = 3.85). Gender comparisons used two-tailed Welch's independent-samples t-tests. After Bonferroni adjustment across the five items in this use-case family, none of the differences between women and men was statistically significant (adjusted p = 0.367 for Q5 and 1.000 for Q1-Q4; raw p-values ranged from 0.073 to 0.672).

*Benefits of the use case.* Regarding the benefits of this use case (Table 14), only 21 participants (7.1% of the sample) reported no benefits. The rest have reported at least one benefit. They highlighted the possibility of viewing certain products of interest (their 3D models) from various

[7] https://www.ikea.com/es/en/customer-service/mobile-apps/ (last access: September 2026)

perspectives or points of view (67.9%), allowing a more attractive and personalized experience (66.6%), promoting social integration (45.7%), as well as providing greater knowledge/information about the main TV content (44.4%).

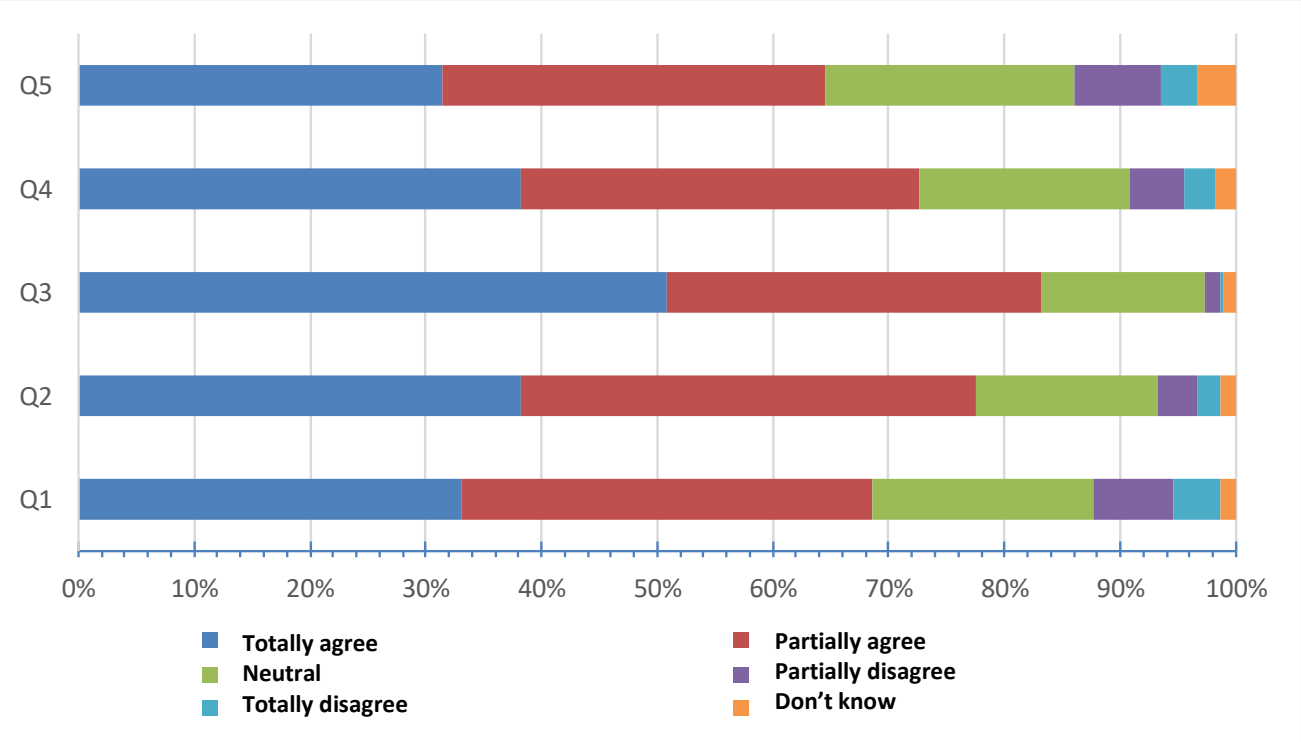

Figure 11. Participants' answers about statements regarding use case 4

Table 10. Statistical values of the responses about statements regarding use case 4

| Statement | Mean value | Standard Deviation |
|---|---|---|
| **Q1** | 3.88 | ±1.081 |
| **Q2** | 4.10 | ±0.930 |
| **Q3** | 4.33 | ±0.795 |
| **Q4** | 4.02 | ±1.010 |
| **Q5** | 3.85 | ±1.062 |

*Limitations of the use case.* Regarding the weaknesses of this use case (Table 15), 35 participants (12%) reported no limitations. The rest declared one or more limitations, highlighting technical inconveniences, such as when the Internet connection is not working or is not fast enough to download supplementary content on time (51.9%); the cost of secondary devices (50.2%); and the distraction (34.1%) or inconvenience of watching TV-related content on secondary devices (29.7%). It is also worth noting that 76 participants (25.9%) highlighted a lack of technical skills or training to use the secondary devices and the service as a limitation in this use case; 56 participants (19.1%) pointed out the potential problems associated with using battery-powered devices; 44 participants (15%) believe the technology is not yet mature enough; and 43 participants (14.7%) foresee potential visual fatigue or harmful effects.

*Participants' interest in this use case.* Participants were asked whether, if this service/use case were available on their home TV, they had the necessary knowledge and the appropriate devices, they would be interested in an experience like the one described in the use case. 205 participants (69.97% of the sample) responded affirmatively (Table 4), while 53 responded negatively (18.09%). Participants who answered affirmatively were asked (open-ended question) to indicate in what situations they considered an experience like the one in the use case useful and in what specific scenarios or applications they would use it. They would mostly use it for viewing advertised products on TV in more detail. Those participants were also asked to indicate the TV programs or genres in which they think supplementary 3D content would be most useful. The responses are summarized in Table 11, which, contrarily, shows that they would find it most useful while watching movies (55.6%), series (45.4%), or educational content (31.7%). As shown in the table, as in use case 3, men cited sports programs more frequently, the only statistically significant difference between the two genders (z-tests with Bonferroni correction).

If their costs were not a problem for participants (they were asked not to consider them), their preferred secondary devices for this use case were VR HMDs (68.3%), AR HMDs (54.6%), smartphones (51.7%), laptops (37.1%) and tablets (31.79%).

Regarding participants' suggestions about this use case, 6 participants (2%) pointed out that the inclusion of sensory effects (e.g., vibration, scent, tactile effects, etc.) would be interesting and would enhance the experience; and 1 participant suggested that it would be interesting that several users could interact with the 3D object simultaneously. Some participants were also concerned about possible visual harm from prolonged use of AR or VR HMDs.

In summary, participants valued the visualization of spatial elements (e.g., 3D models) and enhanced understanding in explanatory contexts. Although approximately 70% of users expressed interest in this use case, perceived complexity and potential visual fatigue or adverse effects were recurring concerns.

Table 11. TV genres for which participants believed having supplementary 3D content would be most useful.

| Content genre | Responses by gender** | | Responses* |
|---|---|---|---|
| | Women | Men | Total |
| **Films** | 64 (61.0%) | 50 (50.0%) | 114 (55.6%) |
| **Series** | 51 (48.6%) | 42 (42.0%) | 93 (45.4%) |
| **Education** | 30 (28.6%) | 35 (35.0%) | 65 (31.7%) |
| **All types of content** | 23 (21.9%) | 31 (31.0%) | 54 (26.3%) |
| **TV reports** | 25 (23.8%) | 26 (26.0%) | 51 (24.9%) |
| **Sports** | 19 (18.1%) | 32 (32.0%) | 51 (24.9%) |
| **Cultural programs** | 25 (23.8%) | 21 (21.0%) | 46 (22.4%) |
| **Musical Videos** | 26 (24.8%) | 19 (19.0%) | 45 (22.0%) |
| **Games/quizzes** | 18 (17.1%) | 20 (20.0%) | 38 (18.5%) |
| **News** | 10 (9.5%) | 15 (15.0%) | 25 (12.2%) |
| **Reality shows** | 11 (10.5%) | 8 (8.0%) | 19 (9.3%) |
| **Events (galas…)** | 10 (9.5%) | 6 (6.0%) | 16 (7.8%) |
| **Music (with lyrics)** | 9 (8.6%) | 6 (6.0%) | 15 (7.3%) |
| **No fiction** | 2 (1.9%) | 5 (5.0%) | 7 (3.4%) |
| **Advertisements** | 3 (2.9%) | 2 (2.0%) | 5 (2.4%) |
| **Documentaries** | 1 (1.0%) | 0 (0.0%) | 1 (0.5%) |
| **Videogames** | 0 (0.0%) | 1 (1.0%) | 1 (0.5%) |
| **N/A** | 3 (2.9%) | 1 (1.0%) | 4 (2.0%) |
| **Total.-** | **105** | **100** | **205 (100%)** |

* % regarding 205 participants
** regarding the number of women and men (105 and 100, respectively).

*B5) Use Case 5 – Combination of traditional (broadcast) TV content with complementary (broadband) 360° video.*

This use case allows users to explore immersive scenes related to the broadcast main TV content. Users can access complementary additional 360º content and watch it on a VR HMD synchronized with the broadcast 2D AV media content watched on the main TV. The sample size for this use case was 234 participants (Table 1).

*Participants' previous Experience.* Only a very small percentage (36 participants, 15.38%) reported prior experience similar to that in this use case (Table 4). Most of the reported experiences involve playing video games (19.4%), watching 360º videos (16.7%), taking 360º virtual tours (11.1%), and using virtual museum apps (8.3%). Only 1 participant reported a negative experience due to distractions from the main content. 4 participants reported experiencing technical problems related to poor Internet connection, operational issues, poor image quality, difficulty of use, and discomfort due to wearing the VR HMDs.

*Perceived Usefulness and Social Value.* Participants provided their level of agreement with the following statements regarding this use case:

*Q1. In general, I find the service/functionality described in this use case useful.*

*Q2. The service/functionality described in this use case provides added value to TV content consumption.*

*Q3. The integration, synchronization, and personalization of the playout of additional 360º video content that complements the content being played on the main TV is a useful and interesting feature.*

*Q4. The service/functionality described in this use case can help increase user attention while consuming TV content.*

*Q5. The service/functionality described in the use case provides a better sense of immersion when consuming 360º video content, complementing the main TV content.*

*Q6. The use case described increases engagement and makes the main TV's content more attractive.*

*Q7. Suitable technology for displaying additional 360º video content, complementary to the main TV content, on secondary devices (e.g., tablets, smartphones, laptops, or VR HMDs) solves a real problem.*

Figure 12 shows the distribution of responses for each statement, and *Benefits of the use case*. Regarding the benefits of this use case (Table 14), only 18 participants (7.7% of the sample) indicated that they did not see any benefit to the use case. The rest have selected or contributed at least one benefit. The main benefits declared were: a more attractive and personalized experience (70.5%); it allows to see other interesting points of view of the scenes or events (63.2%); access to complementary 360º video content that can hook (i.e., engage) viewers to continue watching the main content of the tuned TV channel (47.4%); it makes the viewer feel like a filmmaker, selecting the points of view of the scenes or events that are most interesting in a personalized way (43.6%); it allows for social integration (40.2%).

Table 12 summarizes their statistical values. The sample generally agreed strongly with the statements, and the strongest agreement concerned the sense of presence or being inside the scene (Q5, M = 4.19). Gender comparisons–used two-tailed Welch's independent-samples t-tests. Bonferroni adjustment was applied across the seven conceptually related items in this use-case family. Two differences remained statistically significant after adjustment. Men rated the general usefulness of the use case more highly than women (Q1: women M = 3.65, men M = 4.05), Welch's t (223.74) = -2.88, raw p = 0.0044, adjusted p = 0.0307, Hedges' g = 0.37. Men also rated the integration, synchronization, and personalization of complementary 360° video more highly (Q3: women M = 3.94, men M = 4.33), Welch's t (223.54) = -3.13, raw p = 0.0020, adjusted p = 0.0137, Hedges' g = 0.41. Both effects fall into the small-to-medium range, according to Cohen's conventional benchmarks, suggesting modest gender differences in the perceived usefulness of immersive 360° video integration. The remaining five items were not statistically significant after adjustment. In particular, the difference in whether the technology solves a real problem (Q7: women M = 3.67, men M = 4.03) was nominally significant before correction (raw p = 0.0082) but not after Bonferroni adjustment (adjusted p = 0.0575). Thus, the evidence indicates limited and item-specific gender differences rather than a uniform gender effect across the use case.

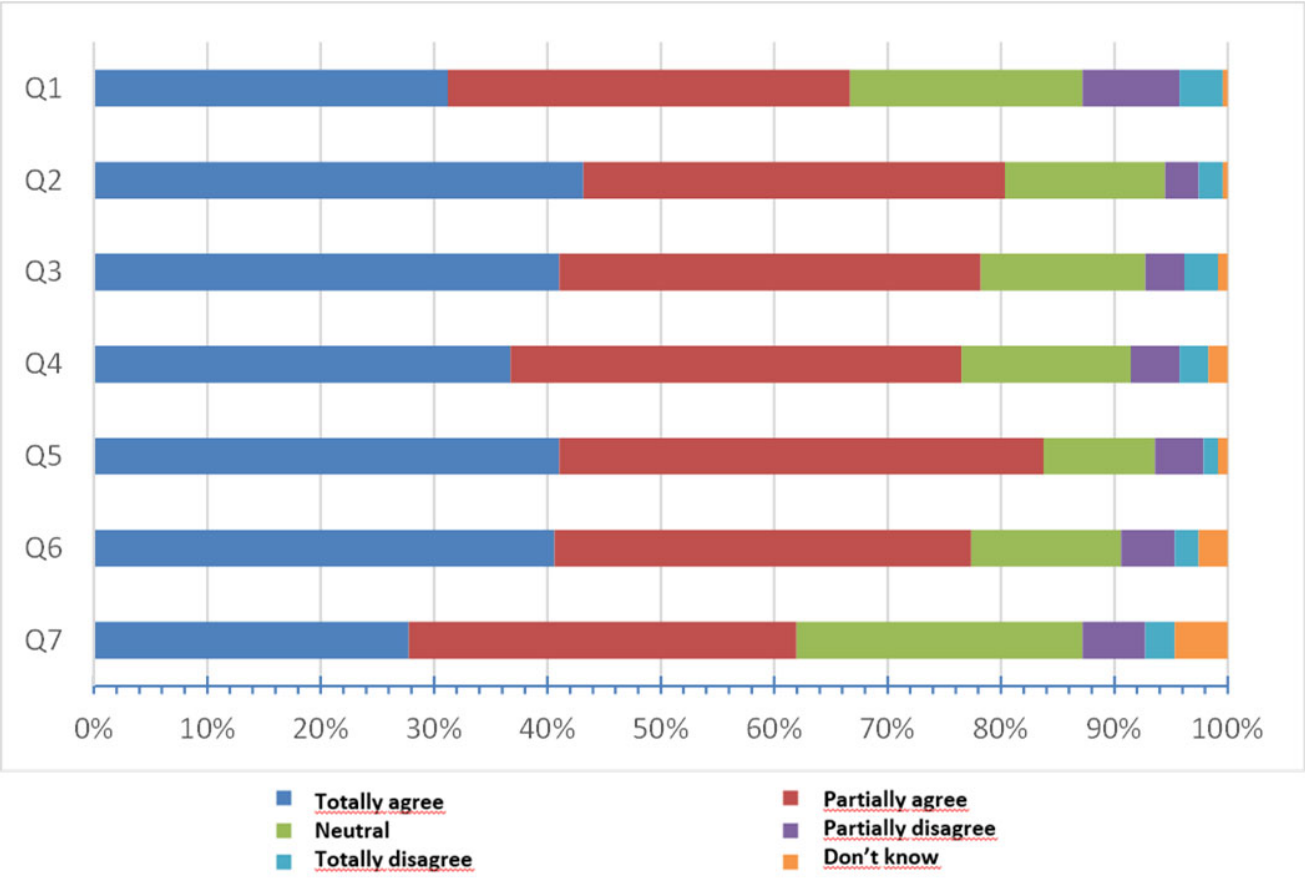

Figure 12. Participants' answers about statements regarding use case 5

Table 12. Statistical values of the responses about statements regarding use case 5

| Statement | Mean value | Standard Deviation |
|---|---|---|
| **Q1** | 3.82 | ±1.088 |
| **Q2** | 4.17 | ±0.930 |
| **Q3** | 4.11 | ±0.981 |
| **Q4** | 4.06 | ±0.967 |
| **Q5** | 4.19 | ±0.877 |
| **Q6** | 4.12 | ±0.966 |
| **Q7** | 3.83 | ±1.003 |

*Limitations of the use case.* Regarding the weaknesses of this use case (Table 15), 36 participants (15.4%) reported seeing no limitations. The rest have declared at least one limitation, highlighting technical inconveniences, such as, for example, when the Internet connection does not work or is not fast enough and the complementary content is not downloaded on time (57.7%) or those related to the lack of synchronization between the playout of the complementary 360º content and that of the main content on the TV (49.1%); the cost of the secondary devices (42.3%), and the discomfort (26.9%) or distraction (23.5%) caused by viewing complementary content on secondary devices. It should be noted that 36 participants (15.4%) pointed out the lack of technological training as a limitation in this use case; 36 participants (15.4%) see possible visual harmful effects; 35 participants (15%) highlighted the problem that using battery-powered devices can pose and the problems that this entails; and 28 participants (12%) believe that the technology is not yet mature enough.

*Participants' interest in this use case.* Participants were asked whether, if this service/use case were available on their home TV, they had the necessary knowledge and the appropriate devices, they would be interested in an experience like the one described in the use case. 156 participants (66.67% of the sample) responded affirmatively (Table 4). Those participants were also asked to indicate the TV programs or genres for which they believed supplementary video content would be most useful. The responses are presented in Table 13. The responses have changed considerably compared to those obtained in the previous study in [3]. Participants would find it most useful when watching films (56.4%), sports (51.9%), or series (45.5%). As already reported in [3], men continue to cite sports programs more frequently. However, the differences between men's and women's responses are not statistically significant (z-tests with Bonferroni correction).

If their costs were not a problem for participants (they were asked not to consider them), their preferred secondary devices for this use case were VR HMDs (71.2%), smartphones (48.1%), laptops (39.7%) and tablets (33.3%).

Participants have provided very few suggestions regarding this use case. 7 participants (3%) proposed including multisensory effects in the experience, 5 (2%) indicated that interaction with others during the experience would be appreciated, and 1 participant mentioned the need for natural interaction (e.g., via voice or gestures) in this type of service.

In summary, although more than 60% of participants showed active interest in this use case as an augmented TV service, it was valued primarily as optional and complementary rather than as a replacement for traditional TV viewing.

Table 13. TV genres for which participants believed supplementary 360º video content would be most useful (multiple choice question)

| Content genre | Responses by gender** | | Responses* |
|---|---|---|---|
| | Women | Men | Total |
| **Films** | 51 (60.0%) | 37 (52.1%) | 88 (56.4%) |
| **Sports** | 40 (47.1%) | 41 (57.7%) | 81 (51.9%) |
| **Series** | 42 (49.4%) | 29 (40.8%) | 71 (45.5%) |
| **Musical Videos** | 23 (27.1%) | 16 (22.5%) | 39 (25.0%) |
| **Games/quizzes** | 23 (27.1%) | 13 (18.3%) | 36 (23.1%) |
| **All types of content** | 16 (18.8%) | 19 (26.8%) | 35 (22.4%) |
| **Cultural programs** | 19 (22.4%) | 16 (22.5%) | 35 (22.4%) |
| **TV reports** | 20 (23.5%) | 14 (19.7%) | 34 (21.8%) |
| **Education** | 19 (22.4%) | 14 (19.7%) | 33 (21.2%) |
| **Events (galas…)** | 21 (24.7%) | 7 (9.9%) | 28 (17.9%) |
| **Reality shows** | 11 (12.9%) | 9 (12.7%) | 20 (12.8%) |
| **Music (with lyrics)** | 8 (9.4%) | 3 (4.2%) | 11 (7.1%) |
| **News** | 5 (5.9%) | 4 (5.6%) | 9 (5.8%) |
| **No fiction** | 1 (1.2%) | 1 (1.4%) | 2 (1.3%) |
| **Documentaries** | 1 (1.2%) | 1 (1.4%) | 2 (1.3%) |
| **Marketing channels** | 1 (1.4%) | 0 (0.0%) | 1 (0.6%) |
| **Total.-** | 85 | 71 | 156 (100%) |

* % regarding 156 participants
** regarding the number of women and men (85 and 71, respectively).

## VII. Discussion

Connected TV penetration is very high in Spain, growing from 82% in 2022 to 95% in 2026 [10], representing approximately 34.3 million individuals. The advent of hybrid media delivery technologies combined with second-screen applications opens the door to innovative future Augmented Hybrid TV services. The findings from the presented study should be interpreted as evidence of perceived usefulness, stated interest, and anticipated user requirements, rather than of observed adoption or technically validated QoE.

When evaluating consumer responses across the five evaluated use cases, several overarching patterns emerge:

- *Limited prior familiarity*: Baseline user experience remained low across the ecosystem, with the exception of Use Case 1 (38.8% prior experience with subtitle customization via OTT, albeit without AR overlays).
- *Immediate potential for accessibility services*: Accessibility-driven features demonstrated the highest immediate adoption potential, with adaptive subtitles receiving consistent positive ratings across perceived usefulness, social inclusion, and practicality.
- *Conditional acceptance of immersive augmentation*: While XR enhancements elicited strong consumer interest (> 64%), acceptance remains contingent upon device ergonomics, sub-second synchronization, seamless domestic integration and potential visual fatigue or adverse effects.
- *Multi-device synchronization as a critical enabler*: Consistent with prior research [3], consumer interest is not the primary bottleneck; technological maturity and inter-device synchronization serve as decisive determinants.
- *Cognitive load and distraction risks*: Complementary secondary-screen content introduces potential risks of cognitive overload if presentation dynamics are not carefully orchestrated.
- *Predominantly similar, but not fully homogeneous, gender usefulness perceptions*: After applying two-tailed Welch's independent-samples t-tests and Bonferroni adjustment separately within each conceptually related item family, most comparisons did not show statistically significant gender differences. Nevertheless, complete homogeneity cannot be claimed. Men reported higher agreement on one item in Use Case 1, concerning whether the subtitle technology solves a real problem, and on two items in Use Case 5, concerning general usefulness and the usefulness of integrating, synchronizing, and personalizing complementary 360° video. The effects that remained statistically significant after adjustment fell within the small-to-medium range (Hedges' g = 0.37-0.41), indicating modest, item-specific gender differences rather than broad gender divergence.

Table 14. Identified benefits for each use case

| Use Case | Benefits | Answers (%) |
|---|---|---|
| 1. Adaptive and Personalized Subtitles. | It can contribute to social integration. | 77 |
| | It allows enjoying a more attractive and personalized experience. | 76 |
| | Useful for learning other languages. | 74 |
| | It allows watching TV without sound in the room. | 59.5 |
| | I do not see any benefit. | 0.8 |
| 2. Complementary (broadband) video content displayed on AR HMD, providing additional information synchronized with the main (broadcast) TV program. | It can contribute to social integration. | 70.6 |
| | It allows enjoying a more attractive and personalized experience. | 65.5 |
| | Useful for learning other languages. | 52.1 |
| | Access to complementary video content can hook viewers (i.e., engage them) to continue watching the main content of the tuned TV channel. | 46.6 |
| | Greater knowledge/information about the main broadcast content watched on the main TV device. | 44.1 |
| | I do not see any benefit. | 4.6 |
| 3. Virtual Large Screen via AR HMDs. | Enhanced visual experience due to a better visualization of contents on a large-scale display. | 70.2 |
| | It can contribute to social integration. | 59.2 |
| | It allows enjoying a more attractive and personalized experience. | 56.3 |
| | Improved viewing conditions in constrained spaces, it allows large-screen viewing regardless of the physical size of the actual TV and the room | 53.9 |
| | I do not see any benefit. | 6.5 |
| 4. Combination of traditional (broadcast) 2D TV content with synchronized virtual 3D objects/models/animations presented in AR/VR HMD devices, enabling spatial visualization of complementary information. | It allows viewing of certain products of interest (their 3D models) from various perspectives. | 67.9 |
| | It allows enjoying a more attractive and personalized experience. | 66.6 |
| | It can contribute to social integration. | 45.7 |
| | It can provide greater knowledge/information about the main TV content. | 44.4 |
| | I do not see any benefit. | 7.1 |
| 5. Combination of traditional (broadcast) TV content with complementary (broadband) 360° video. | It allows enjoying a more attractive and personalized experience. | 70.5 |
| | It allows seeing other interesting points of view of the scenes or events | 63.2 |
| | Access to complementary 360º video content can hook (i.e., engage) viewers, encouraging them to continue watching the main content of the tuned TV channel. | 47.4 |
| | It can make the viewer feel like a filmmaker, selecting the points of view of the scenes or events that are most interesting, in a personalized way. | 43.6 |
| | It can contribute to social integration. | 40.2 |
| | I do not see any benefit. | 7.7 |

Table 15. Identified limitations for each use case

| Use Case | Limitations | Answers (%) |
|---|---|---|
| 1. Adaptive and Personalized Subtitles. | The lack of synchronization of the subtitles with the AV content. | 57.4 |
| | The subtitles could not be downloaded on time due to technical issues (e.g., a poor Internet connection). | 55 |
| | Distraction or lack of concentration on the AV content. | 34.7 |
| | Discomfort or annoyance caused by reading the subtitles in a virtual space outside the TV frame (when using AR HMDs). | 59.5 |
| | The cost of these secondary devices. | 31.8 |
| | The inconvenience of having to read the subtitles on secondary devices (when used). | 29.3 |
| | Problems that battery-powered devices can pose and other problems they entail. | 15.7 |
| | Lack of technical skills or training to use the service. | 14 |
| | Possible visual fatigue or harmful effects. | 9.1 |
| | I do not see any limitation. | 13.6 |
| 2. Complementary (broadband) video content displayed on AR HMD, providing additional information synchronized with the main (broadcast) TV program. | The complementary content could not be downloaded on time due to technical issues (e.g., a poor Internet connection). | 60.9 |
| | The lack of synchronization of complementary content with the AV content. | 59.7 |
| | The cost of these secondary devices. | 48.7 |
| | Discomfort or annoyance caused by visualizing complementary content on secondary devices (tablets, smartphones, laptops, AR/VR HMDs, etc.). | 36.1 |
| | Distraction or lack of concentration on the AV content. | 32.4 |
| | Problems that battery-powered devices can pose and other problems they entail. | 22.3 |
| | Lack of technical skills or training to use the service. | 21.4 |
| | Possible visual fatigue or harmful effects. | 18.9 |
| | Technology is not yet mature enough. | 16.4 |
| | I do not see any limitation. | 7.1 |
| 3. Virtual Large Screen via AR HMDs. | The cost of these secondary devices (AR HMDs). | 57.6 |
| | In shared consumption sessions, interacting with more people watching TV in the same room, the lack of synchronization of the virtual (AR-based) TV content with the actual TV content. | 49.4 |
| | Discomfort or annoyance caused by visualizing TV content on the virtual TV using AR HMDs. | 40 |
| | Lack of technical skills or training to use the AR HMDs and the service. | 27.3 |
| | Problems that battery-powered devices can pose and other problems they entail. | 24.9 |
| | Possible visual fatigue or harmful effects. | 22.9 |
| | Technology is not yet mature enough. | 18 |
| | I do not see any limitation. | 10.2 |
| 4. Combination of traditional (broadcast) 2D TV content with synchronized virtual 3D objects/models/animations presented in secondary devices, enabling spatial visualization of complementary information. | The complementary content could not be downloaded on time due to technical issues (e.g., a poor Internet connection). | 51.9 |
| | The cost of these secondary devices (AR HMDs). | 50.2 |
| | Distraction or lack of concentration on the AV content. | 34.1 |
| | Discomfort or annoyance caused by visualizing complementary content on secondary devices. | 29.7 |
| | Lack of technical skills or training to use the secondary devices and the service. | 25.9 |
| | Problems that battery-powered devices can pose and other problems they entail. | 19.1 |
| | Technology is not yet mature enough. | 15.0 |
| | Possible visual fatigue or harmful effects. | 14.7 |
| | I do not see any limitation. | 12 |
| 5. Combination of traditional (broadcast) TV content with complementary (broadband) 360° video. | The complementary 360º content could not be downloaded on time due to technical issues (e.g., a poor Internet connection). | 57.7 |
| | In shared consumption sessions, interacting with more people watching TV in the same room, the lack of synchronization between the complementary 360° video content and the main TV content. | 49.1 |
| | The cost of these secondary devices (VR HMDs). | 42.3 |
| | Discomfort or annoyance caused by visualizing complementary content on secondary devices. | 26.9 |
| | Distraction or lack of concentration on the AV content. | 23.5 |
| | Lack of technical skills or training to use the secondary devices and the service. | 15.4 |
| | Possible visual fatigue or harmful effects. | 15.4 |
| | Problems that battery-powered devices can pose and other problems they entail. | 15.0 |
| | Technology is not yet mature enough. | 12.0 |
| | I do not see any limitation. | 15.4 |

These cross-case findings outline a dual trajectory for Augmented Hybrid TV services: accessibility-enhancing features are mature and immediately deployable, whereas immersive AR/VR extensions are promising but still require development in ergonomics and technology. The following subsections discuss these dimension dynamics in detail.

### *A. Interest in Augmented Hybrid TV services for different content genres*

Leisure apps (e.g., sports, games, films, reality shows, or music) are expected to drive content production for hybrid and shared TV media services. As expected, men were significantly more interested in extra-related sports content than women. Unlike the results of our previous study in [3], interest in extra-related content in entertainment genres, such as films, sports, or series, overtook interest in non-entertainment genres, such as news, non-fiction, or documentaries. This new study reveals that other genres, such as musical videos, games, and cultural and educational programs, are also of interest to consumers. So, they should also be considered by content producers and providers when offering those services.

### *B. Analyzed use cases*

As of the date the survey took place, the authors did not identify any commercially deployed services that exactly matched the five evaluated configurations. As expected, aside from use case 1, very few participants had previously participated in similar experiences (Table 4). 38.9%, 2.1%, 0%, 14%, and 15.38% of participants in each sample reported having had prior experiences similar to use cases 1 to 5, respectively. Nevertheless, most of the participants consider that the service/functionality described in each use case provides added value to TV content consumption experience and showed interest in all the use cases (76.86%, 64.71%, 66.94%, 69.97% and 66.67%, respectively, as shown in Table 4). Participants identified clear benefits for all, but some also think there are still serious limitations to overcome and that technology needs further enhancement to successfully deploy those services (Table 15). Synchronization, device ergonomics, connectivity, and usability were frequently identified as anticipated limitations. Notice that there is very low penetration of basic hybrid TV media services with second-screen apps in Spain (even in Europe), and, to the authors' knowledge, no commercial hybrid TV services offering broadband-complementary AR/VR content are available yet.

One of the main identified technical limitations is the lack of synchronization between broadcast and broadband content playout. Given the rising research interest in media synchronization in hybrid TV services and the disadvantages of proprietary technologies (e.g., fingerprinting or watermarking techniques [29], [30]), researchers, broadcasters, telecommunications operators and manufacturers have started developing new international standards ([31], [32]). Nevertheless, more work is still needed in that direction. From version 2.0.1 of the HbbTV standard, IDES functionalities and guidelines for achieving hybrid media delivery and synchronous consumption of broadcast and related broadband content were already specified. However, the current version of the standard does not yet provide full-fledged media synchronization solutions.

Regarding the second main technical limitation identified, bandwidth insufficiency, it can hinder the seamless delivery of broadband content and interactive features. Nevertheless, it can be overcome by improving communication networks, infrastructure, and technologies. In Spain, the major operators are permanently increasing the deployment of their fiber optic networks and improving their wireless access networks (e.g., with 5G technology). Despite these advances, the rise in demand for high-quality (e.g., UHD and SUHD) media content will require additional network resources (e.g., bandwidth), which, in turn, will necessitate greater effort and improved solutions to address potential congestion and scalability issues.

It is also worth noting that varying levels of technological literacy among viewers will affect engagement with complementary content control and with interactive features. The digital divide creates disparities in access to Augmented Hybrid TV services.

Accessibility-oriented services received the most favorable evaluations. Among the evaluated use cases, the adaptive subtitle service clearly stands out as the most readily understood, positively evaluated and immediately applicable. Participants strongly agreed on its usefulness, particularly regarding social inclusion and accessibility. The highest-rated statements highlighted the role of customizable subtitles in facilitating integration for people with AV impairments, enabling language accessibility for non-native speakers, and supporting language learning.

These results highlight the relevance of accessibility-driven services in the evolution of Augmented Hybrid TV platforms. While immersive experiences often attract attention for their novelty, the study's results suggest that accessibility-oriented features may represent the most realistic short-term path for innovation. In this context, personalization capabilities, language selection, and the possibility of presenting subtitles on secondary devices can significantly improve the usability of hybrid TV services in heterogeneous viewing environments.

Furthermore, the widespread use of second-screen devices identified in the survey suggests that multi-device subtitle delivery could be implemented without requiring additional hardware adoption. This aligns with previous research on hybrid TV systems, where accessibility and personalization have been identified as key drivers of user acceptance.

### *C. Immersive augmentation generates high interest but conditional acceptance*

In contrast with the interest in accessibility features, immersive TV scenarios involving AR/VR technologies generated strong conceptual interest but more cautious evaluations of practical adoption. Participants recognized the potential value of complementary AR content, 3D objects, and immersive experiences in enhancing enjoyment and engagement with AV content. However, these benefits were accompanied by concerns related to usability, ergonomics, and technological maturity. Moreover, adoption will require seamless integration into existing domestic TV and multi-device environments. In this context, seamless integration refers to interoperability among the main TV, companion devices, AR/VR HMDs, and broadcast/broadband services, including the straightforward discovery, configuration, control, and synchronized presentation of related content without disrupting conventional TV viewing. Several factors were identified by participants as potential reasons for this conditional interest. First, the relatively low level of familiarity with AR/VR technologies observed among participants suggests that augmented immersive TV services are still perceived as exploratory. Although a significant proportion of participants had previously experienced AR or VR, these experiences were typically limited to entertainment contexts such as gaming or short demonstrations. Second, participants expressed concerns regarding physical comfort, visual fatigue, and the potential isolation from co-viewers when using immersive devices. TV consumption has traditionally been a social activity, and immersive technologies that require individual HMDs may disrupt shared viewing practices.

The findings support the evaluation of immersive services as optional extensions rather than replacements for conventional TV viewing. Hybrid configurations that combine traditional TV screens with complementary immersive AR/VR content appear promising.

### *D. Multi-device ecosystems as the enabling environment for augmented TV*

The survey confirms that multi-device environments are already a common characteristic of contemporary TV consumption. A large majority of participants reported using second-screen devices while watching TV, with smartphones being the dominant companion device. This behavior provides an important foundation for Augmented Hybrid TV services. Rather than requiring disruptive changes in viewing habits, many augmentation features could be implemented using devices already present in the user's environment. For instance, complementary information, subtitle personalization, or interactive features can be delivered through mobile devices synchronized with the main TV content. However, the results also highlight the importance of accurate synchronization and seamless interaction between devices. Our previous research on hybrid TV systems in [3] already identified synchronization as a key technical challenge affecting the user’s QoE. The present study reinforces this observation, indicating that synchronization failures were consistently identified by participants as an anticipated technical limitation. Consequently, robust synchronization mechanisms and low-latency content delivery should be treated as priority requirements in the development of augmented hybrid TV platforms and in QoE evaluation.

### *E. Design implications for future broadcast and Hybrid TV Systems*

The findings of this study provide several implications for the design of future augmented hybrid TV platforms and systems. First, accessibility and personalization features should be prioritized as core components of Augmented Hybrid TV services. Adaptive subtitles, language customization, and multi-device presentation can significantly enhance the inclusiveness of TV experiences. Second, immersive technologies should be integrated progressively, prioritizing complementary and optional features that enhance specific content types, such as sports events, documentaries, and educational programs. Third, system architectures must support multi-device synchronization and flexible content distribution. The relationships between these three design dimensions and their empirical grounding are summarized in Table 16. As viewers increasingly consume TV content in heterogeneous environments involving multiple connected devices, future broadcast systems must be designed to coordinate AV streams across those devices while maintaining a consistent user experience. Table 16 synthesizes the preceding discussion by mapping the key empirical findings onto the design priorities that participants implicitly or explicitly identified, together with their indicative implications for hybrid TV system design. It should be noted that the table reflects user-perceived needs and anticipated requirements as derived from a stated-preference survey; it does not constitute a technical validation of these requirements, nor does it prescribe specific implementation solutions. Rather, it is intended to provide researchers, broadcasters, and system developers with a structured reference on the user perspective, which should inform, but not substitute for, technical design and standardization decisions.

The priorities identified in Table 16 not only have implications for system design but also open specific opportunities for researchers, developers, and market stakeholders, as discussed in the following subsection.

Table 16. Summary of empirical evidence, user-identified design priorities, and implications for hybrid TV system design

| Empirical evidence | User-identified design priority | Indicative implication for Hybrid TV Systems |
|---|---|---|
| High perceived usefulness of adaptive and personalized subtitles | Individual accessibility and personalization | Broadband delivery of personalized accessibility components (e.g., subtitles or sign language video) |
| Synchronization consistently rated as a critical limitation across all use cases | Temporal alignment between hybrid (i.e., broadcast and broadband) streams | Inter-stream and inter-device synchronization mechanisms as a baseline requirement |
| Smartphones as the dominant companion device | Compatibility with widely available consumer devices | Lightweight discovery, pairing, and session-management protocols for mobile devices |
| Strong interest in AR/VR, but conditional on usability and comfort | Progressive and optional augmentation | Conventional-TV fallback mode; immersive features as opt-in extensions |
| Cost and ergonomic concerns were identified across all use cases | Device flexibility and reduced hardware dependency | Device-agnostic service architecture supporting heterogeneous terminal types |
| Distraction and cognitive load were raised as anticipated limitations | User-controllable information presentation | Optional activation, adjustable information density, and user-driven filtering |

### *F. Opportunities for media researchers, developers and market stakeholders*

The HbbTV standard has marked a significant milestone in promoting the adoption of hybrid TV media services, mostly in Europe[8]. However, it does not yet offer fully-fledged solutions for the successful deployment of such services. Among the issues that remain to be solved are: (1) the definition of signaling mechanisms capable of discovering, associating, and describing broadband-related media contents; (2) the development of interaction and negotiation procedures among the devices involved; and (3) the provision of adaptive media synchronization approaches, encompassing protocols, algorithms, and playout adjustment methods, to ensure accurate temporal alignment (i.e., synchronization) during the consumption of related hybrid contents.

The primary objective is to enable smooth convergence, coordination, and interoperability among the various broadcast and broadband delivery technologies. By leveraging their respective strengths and complementary capabilities, these technologies should no longer be viewed as separate domains but rather as components of a unified hybrid TV ecosystem. Furthermore, it is desirable to provide transparent interaction and coordination among the various connected devices involved in hybrid TV consumption. Such an approach would broaden the ways in which hybrid content can be accessed and experienced, enabling future experimentation with services and business models.

Many participants agreed that technology is not yet mature enough to offer these types of services with sufficient guarantees, that some technical background or skills are needed, and that they are concerned about the cost of the devices, about problems related to the use of battery-powered devices, and about possible visual fatigue or harmful effects of wearing HMDs during prolonged use (see Table 15). Additional research efforts are needed to address the identified limitations, including the development of emerging technologies, the evolution of standards, improved media synchronization mechanisms, more advanced applications, and the availability of more affordable and user-friendly devices. Overcoming these challenges will help enhance users' QoE, thereby increasing the likelihood of widespread acceptance and successful deployment of Augmented Hybrid TV services.

The growing multi-screen ecosystem offers many opportunities for developers to design novel TV-related services, including second-screen applications that complement TV programs and multi-screen solutions that enable media consumption across a wide range of devices and displays. By operating in an integrated manner, these applications can provide a unified user experience while addressing interoperability issues. Nevertheless, their design must be approached carefully to avoid being perceived as disruptive, overwhelming, or intrusive. In addition, developers should continue incorporating new functionalities that reflect the ongoing evolution of hybrid TV practices, particularly considering that broadcasters are increasingly leveraging these behaviors for purposes such as advertising and audience engagement. New commercial opportunities related to hybrid TV services will emerge. For example, advertisers can leverage second-screen engagement for synchronized advertising campaigns. They can launch real-time marketing opportunities during TV events. Analytics from users' and second-screen interactions can be collected, providing valuable data on viewer behavior and preferences. This way, targeted advertising strategies could also be defined.

Finally, some studies have focused on determining the acceptable asynchrony ranges that are tolerable to human perception regarding different use cases in which media synchronization is needed (see Section III of [3]). Excessive asynchrony could impair perceived QoE and user satisfaction. The allowable limits for the different hybrid TV media services have not yet been rigorously determined. More QoE tests will be needed to determine them. Such limits should be established through extensive research and very rigorous objective and subjective assessments, possibly including longer-term testing in live systems rather than in artificial test environments. Likewise, such experiments should be conducted for each particular augmented hybrid TV service use case under study, as the tolerable range of asynchrony levels strongly depends on the usage scenario. Having accurate references for such limits will be highly relevant to network operators, content providers, and developers, helping ensure they do not exceed them.

[8] https://www.hbbtv.org/deployments/ (last access: September 2026)

Together, these research directions may help improve the Augmented Hybrid TV media experience and the value of related multi-screen, immersive, and interactive applications.

## VIII. Limitations of the Study

The study provides a preliminary assessment of perceived acceptance and does not measure actual system use, willingness to pay, continued-use intention, market adoption, or QoE resulting from interaction with functional prototypes.

The empirical findings presented in this paper are based on a quantitative CAWI survey designed to capture Spanish consumers' preferences, expectations, and perceived utility regarding Augmented Hybrid TV services. While this cross-sectional approach enabled a relatively large nationwide non-probability quota sample ($N = 626$) rather than a small, constrained laboratory cohort, online survey methodologies inherently present potential risks, including self-selection bias, coverage error, and reliance on self-reported perceptions rather than observed behavioral adoption or willingness to pay. To systematically minimize these biases, several active methodological controls were enforced during study design and fieldwork execution.

First, the potential coverage error was bounded by restricting the target universe strictly to active connected TV and Internet users, ensuring alignment between the digital delivery channel and the evaluated ecosystem. Second, selection bias was actively mitigated through stratified proportional quota sampling by sex and age, administered by an independent market research agency (Ikerfel S.A.). Third, to eliminate mechanical, speeder, or inattentive response patterns, embedded attention-check items and duration-screening protocols were implemented, ensuring high data fidelity across the retained sample.

A second limitation concerns the evaluation of emerging, non-commercially deployed technologies. Because fully integrated AR/VR hybrid broadcast systems are not yet commercially available, respondents evaluated standardized, video-based demonstrations of potential use cases rather than hands-on interactive prototypes. To mitigate cognitive overload, evaluation fatigue, and sequential order bias, each participant was exposed to a randomly assigned pair of use cases from the five evaluated scenarios. Consequently, the findings represent stated user openness, perceived utility, and preliminary interest dynamics within the baseline context of Spanish connected TV users. Future research will build upon these empirical baselines by developing functional prototype implementations and conducting Quality of Experience (QoE) assessments in both controlled laboratory and real-world broadcast testing environments.

## IX. Conclusions

An exploratory assessment of user requirements and preliminary perceptions of acceptance across five representative use cases of Augmented Hybrid TV services that integrate AR/VR technologies, multi-device interaction, and accessibility features has been presented in this paper. The findings support continued investigation of hybrid service configurations combining accessibility, personalized interaction, and optional immersive extensions. The results also reveal a clear distinction between accessibility-oriented services and immersive augmentation technologies in terms of perceived maturity and potential for acceptance.

The main conclusions of this work can be summarized as follows:

1. *Empirical evidence on perceived usefulness and stated interest in Augmented Hybrid TV services.* The study provides one of the first large-scale analyses of user opinions on Augmented Hybrid TV experiences, which combine traditional TV with AR/VR technologies and multi-device environments.
2. *Identification of accessibility services as the most readily understood and positively evaluated innovation pathway.* Adaptive subtitle services and AR/VR-based accessibility features were consistently rated as highly useful and socially valuable, highlighting them as potential candidates for earlier implementation and prototype evaluation.
3. *Characterization of anticipated barriers for immersive TV technologies.* Although AR/VR-based services generate strong interest, their acceptance is constrained by usability, comfort, and technological familiarity factors. These findings suggest that immersive extensions should be designed as complementary experiences rather than replacements for conventional TV viewing.
4. *Confirmation of multi-device ecosystems as a key enabler of future hybrid TV services.* The widespread use of second-screen devices indicates that Augmented Hybrid TV services can be integrated into existing user environments if synchronization and usability challenges are adequately addressed.

The findings identify robust, scalable, and standards-compliant delivery, association, and synchronization mechanisms as priorities for subsequent system development and evaluation. The synchronization requirement is particularly relevant for accessibility-oriented and immersive services, for which synchronization failures were consistently identified as a major limitation. Table 16 consolidates these findings into a structured mapping of empirical evidence, user-perceived design priorities, and indicative system-level implications. It should be read as a user-perspective reference derived from stated-preference data, not as a set of technically validated requirements.

The findings provide user-centered evidence that can inform the design and subsequent technical evaluation of Augmented Hybrid TV services. The development of appropriate, standards-compliant technological solutions and comfortable equipment should continue, together with the production of suitable content and the deployment of augmented services and applications. Marketing and commercial initiatives are also advisable to facilitate their adoption. The authors expect the findings of this study to help interested stakeholders understand the remaining challenges, identify emerging opportunities, and prioritize efforts aimed at supporting the adoption of these services.

## Appendix A: Survey Technical Specifications

The following table summarizes the operational metrics and comprehensive structural features of the empirical survey. This datasheet acts as a macro-level reference point for replication frameworks targeting multi-device immersive media consumption.

Table A1. Main technical features of the CAWI survey

| **Parameter** | **Specification Value** |
|---|---|
| Survey Mode / Methodology | Computer-Assisted Web Interviewing (CAWI) |
| Target Population | Spanish connected Internet and TV users aged 18–64 |
| Geographic Scope | Participants recruited across Spain |
| Fieldwork Data Collection Period | November 1 – December 31, 2025 |
| Panel Administration Agency | Ikerfel S.A. |
| Sampling Framework | Non-probability quota sampling, with quotas for gender and age |
| Total Sessions Initiated | 1,197 |
| Valid Responses Retained (N) | 626 |
| Completion-and-retention proportion | 52.3% of initiated sessions |
| Average Completion Time | 27 min 55 s (Range: 23:32 - 35:21) |
| Individual Use Case Sample Limits (n) | n = 234 to n = 293 per individual scenario |

## Acknowledgment

The authors would like to thank their students Rubén Prieto Molina and Juan González Salinas for their help in preparing the survey.

## AUTHORS' BIOGRAPHIES

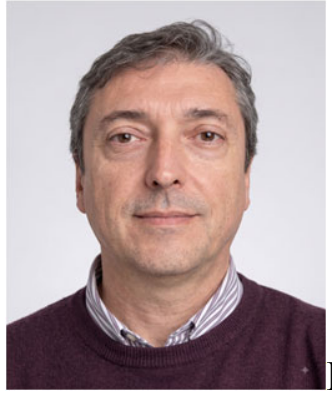

**Fernando Boronat** (M'93–SM'11), was born in Gandia (Spain), and went to the Universitat Politècnica de València (UPV) in Spain, where he studied Telecommunications Engineering. He received the M.E. and PhD degrees in telecommunication engineering from the UPV in 1994 and 2014, respectively. After working for several Spanish telecommunications companies, he returned to the UPV in 1996. He is currently a Full Professor in the Communications Department. He is the Head of the Immersive Interactive Media R&D Group (http://iim.webs.upv.es) at the Gandia Campus of UPV. His main topics of interest are communication networks, multimedia systems, multimedia protocols, and media synchronization. He is the author of two books, several book chapters, an IETF RFC and more than 100 research papers. He is currently involved in several IPCs of national and international journals and conferences. He is a member of IEEE and ACM.

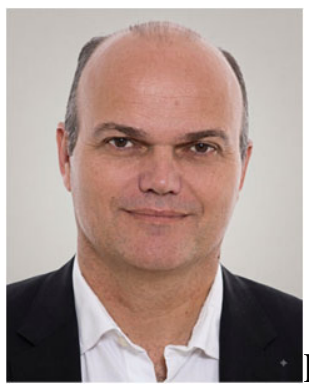

**Bernat Roig-Merino** was born in Valencia, Spain. He received the M.S. degree in Agricultural Engineering and the PhD degree in Economics and Social Sciences from the Universitat Politècnica de València (UPV), Valencia, Spain, in 1993 and 2004, respectively. Since 1998, he has been with the Department of Economics and Social Sciences at UPV, where he is currently an Associate Professor at the Campus de Gandia. He is also a researcher at the Research Institute for Integrated Coastal Zone Management (IGIC) at UPV. His main research specialization lies in social and market research methodologies, with extensive expertise in quantitative survey design, Computer-Assisted Web Interviewing (CAWI), qualitative focus group frameworks, market segmentation, and user acceptance modeling. Over his career, he has led and participated in numerous competitive R&D projects and industry contracts focused on social impact assessment, territorial dynamics, and smart technology adoption, including national projects funded by the Spanish Government (e.g., READI, GO INNOWATER, and HYBSYNC). Dr. Roig-Merino is the author or co-author of over 30 peer-reviewed scientific papers in top-tier journals, book chapters, and international conference proceedings.

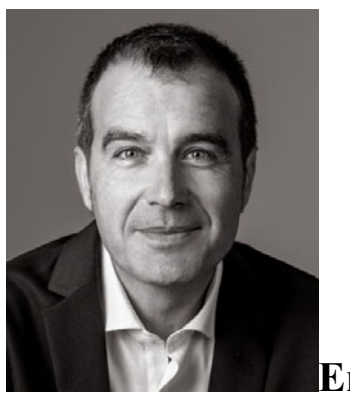

**Enric Sigalat** (Gandia, Spain) holds a PhD in Sociology and degrees in Sociology and Social Work from the University of Valencia (Universitat de València, UV), where he is an Associate Professor in the Department of Social Work and Social Services within the Faculty of Social Sciences. He conducts his research at the UV's Inter-University Institute for Local Development as a member of the Territorial Development Research Group (GRIDET). His research career has focused on the participation in and coordination of competitive research projects and contracts at local, regional, national and European levels, with a strong emphasis on applied social research and knowledge transfer. He has authored more than 160 publications, including articles in indexed scientific journals, book chapters, monographs, and papers presented at national and international conferences. His principal research interests include local development, social research methodology, mixed methods research, inclusive smart cities, and citizen participation. He also serves on the scientific and editorial boards of a range of national and international journals and conferences.

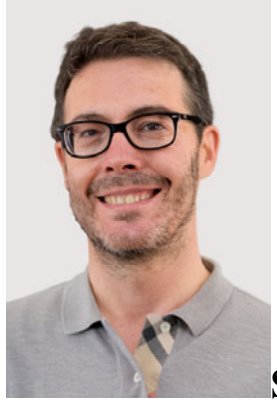

**Santiago J. Flores** received the M.E. degree in telecommunications engineering from the Universitat Politècnica de Catalunya (UPC), Barcelona, Spain, in 1992, and the PhD degree in telecommunications engineering from the Universitat Politècnica de València (UPV), Valencia, Spain, in 1998. He is currently an Associate Professor in the Communications Department at UPV's Gandia Campus and regularly collaborates with the IIM R&D Group. His research interests include multimedia systems and digital signal processing for wireless communications. He has participated in over 20 competitive research projects and has authored 3 research books, 14 educational books, several book chapters, and numerous journal and conference papers in these fields.